\documentclass[12pt,preprint]{aastex}
\usepackage{epsfig}
\usepackage{lscape}
\usepackage{rotating}


\def\ls{{_<\atop^{\sim}}}
\def\gs{{_>\atop^{\sim}}}

\lefthead{Nicastro et al.}

\begin{document}
%%%%%%%%%%%%%%%%%%

\title{{\em Chandra} View of the Warm-Hot IGM toward 1ES~1553+113: Absorption Line Detections and Identifications (Paper I)}

\author{F. Nicastro$^{1,2}$, M. Elvis$^{2}$, Y. Krongold$^3$, S. Mathur$^4$, A. Gupta$^4$, C. Danforth$^5$, X. Barcons$^{6}$, S. Borgani $^{7,8}$, 
E. Branchini$^9$, R. Cen$^{10}$, R. Dav\'e$^{11}$, J. Kaastra$^{12}$, F. Paerels$^{13}$, L. Piro$^{14}$, J.M. Shull$^5$, Y. Takei$^{15}$, L. Zappacosta$^{1}$}

\altaffiltext{1}{Osservatorio Astronomico di Roma - INAF, Via di Frascati 33, 00040, Monte Porzio Catone, RM, Italy}
\altaffiltext{2}{Harvard-Smithsonian Center for Astrophysics, 60 Garden St., MS-04, Cambridge, MA 02138, USA}
\altaffiltext{3}{Instituto de Astronomia, Universidad Nacional Autonoma de Mexico, Mexico City (Mexico)}
\altaffiltext{4}{Ohio State University, Columbus, OH, USA}
\altaffiltext{5}{CASA, Department of Astrophysical and Planetary Sciences, University of Colorado, 389-UCB, Boulder, CO 80309, USA}
\altaffiltext{6}{Instituto de F\'\i sica de Cantabria (CSIC-UC), 39005 Santander, Spain}
\altaffiltext{7}{Dipartimento di Astronomia dell'Università di Trieste, Via G.B. Tiepolo 11, I-34131 Trieste, Italy} 
\altaffiltext{8}{INAF - Osservatorio Astronomico di Trieste, Via G.B. Tiepolo 11, I-34131 Trieste, Italy}
\altaffiltext{9}{Dipartimento di Fisica 'E. Amaldi', Universitá degli Studi 'Roma Tre', via della Vasca Navale 84, 00146 Roma, Italy}
\altaffiltext{10}{Princeton University Observatory, Princeton, NJ 08544, USA}
\altaffiltext{11}{Astronomy Department, University of Arizona, Tucson, AZ 85721, USA}
\altaffiltext{12}{SRON Netherlands Institute for Space Research, Sorbonnelaan 2, 3584 CA, Utrecht, The Netherlands}
\altaffiltext{13}{Columbia Astrophysics Laboratory and Department of Astronomy, Columbia University, 550 West 120th Street, 
New York, NY 10027, USA}
\altaffiltext{14}{INAF-IAPS, Via del Fosso del Cavaliere 100, I-00133 Roma, Italy}
\altaffiltext{15}{Institute of Space and Astronautical Science, Japan Aerospace Exploration Agency, 3-1-1 Yoshinodai, Chuo-ku, Sagamihara, 
Kanagawa 252-5210, Japan}

\begin{abstract}
We present the first results from our pilot 500 ks {\em Chandra}-LETG Large Program observation of the soft 
X-ray brightest source in the $z \gs 0.4$ sky, the blazar 1ES~1553+113, aimed to secure the first uncontroversial  detections of the 
missing baryons in the X-rays. 

We identify a total of 11 possible absorption lines, with single-line statistical significances between 2.2-4.1$\sigma$. 
Six of these lines are detected at high single-line statistical significance ($3.6 \le \sigma \le 4.1$), while the remaining 
five are regarded as marginal detections in association with either other X-ray lines detected at higher significance 
and/or Far-Ultraviolet (FUV) signposts. 
Three of these lines are consistent with metal absorption at $z\simeq 0$, and we identify them with Galactic OI and CII. 
The remaining 8 lines may be imprinted by intervening absorbers and are all consistent with being high-ionization 
counterparts of Far-Ultraviolet (FUV) HI and/or OVI IGM signposts. 

In particular, five of these eight possible intervening absoption lines (single-line statistical significances of 4.1$\sigma$, 4.1$\sigma$, 
3.9$\sigma$, 3.8$\sigma$ and 2.7$\sigma$), are identified as CV and CVI K$\alpha$ absorbers belonging to three WHIM systems at 
$z_X = 0.312$, $z_X = 0.237$ and $\langle z_X \rangle = 0.133$, which also produce broad HI (and OVI for the $z_X = 0.312$ system) 
absorption in the FUV. For two of these systems ($z_X = 0.312$ and 0.237), the {\em Chandra} X-ray data led the a-posteriori discovery 
of physically consistent broad HI associations in the FUV (for the third system the opposite applies), so confirming the power of the 
X-ray-FUV synergy for WHIM studies.  
The true statistical significances of these three X-ray absorption systems, after properly accounting for the number of redshift trials, 
are $5.8\sigma$ ($z_X = 0.312$; 6.3$\sigma$ if the low-significance OV and CV K$\beta$ associations are considered), $3.9\sigma$ 
($z_X = 0.237$), and 3.8$\sigma$ ($\langle z_X \rangle = 0.133$), respectively. 
\end{abstract}


%%%%%%%%%%%%%%%%%%%%%%%%%%%%%%%%%%%%%%%%%%%%%%%%%%%%%%%%%%%%%%%%%%%%%%%%
\section{Introduction}
According to the latest baryon census in the local ($z \ls 0.4$) Universe (Shull, Smith \& Danforth, 2012 and 
references therein), about 30-40\% of the total number of baryons predicted by Big-Bang Nucleosynthesis (e.g. Kirkman et al. 2003), 
inferred by density fluctuations of the Cosmic Microwawe Background (e.g. Bennett et al. 2003; Spergel et al. 2007), and seen at $z\sim 3$ 
in the ``Ly$\alpha$ Forest'' (e.g. Rauch, 1998; Weinberg et al. 1997), are missing at the present epoch. 

Of the low-$z$ baryons we can account for in a recent census (Shull, Smith \& Danforth, 2012), $(18 \pm 4)$\% lie in a collapsed phase 
(stars, cold atomic and molecular gas in galaxies, circumgalactic medium - CGM -, and hot virialized gas in groups and clusters of galaxies), 
$(28 \pm 11)$\%  lie in the local  photoionized  Ly$\alpha$ forest, and finally, as much as $(25 \pm 8)$\% , are estimated to be found in the 
OVI/BLA-traced Warm-Hot Intergalactic Medium
\footnote{This estimate is based on an average ionization and metallicity, versus column density, 
correction derived by hydrodynamical simulations, and therefore suffers not only the large scatter found in the given simulation 
between these parameters (i.e. Fig. 8 in Shull, Smith \& Danforth, 2012), but also the scatter of theoretical predictions from simulation to 
simulations (e.g. Cen \& Ostriker, 2006, Tornatore et al. 2010, Bertone et al. 2010, Branchini et al. 2009, Smith et al. 2011). 
The resulting number of baryons still missing also depends on these uncertainties.}
. The remaining $(29 \pm 13)$\% of baryons still remains elusive, and are likely to lie in the hot phase of the WHIM, at temperatures 
larger than logT$\simeq 5.5$ K. 

This is predicted by nearly all hydrodynamical simulations for the formation of structures in a $\Lambda$-CDM 
Universe (e.g. Cen \& Ostriker, 2006, Tornatore et al. 2010, Bertone et al. 2010, Branchini et al. 2009, Smith et al. 2011). 
According to simulations, starting at redshift of $z\sim 2$ some of the IGM was shock-heated to temperatures of $10^5-10^7$ K 
during the continued process of collapse and structure formation, and enriched up to Z$_{WHIM}=0.1-1$ Z$_{\odot}$ by galaxy super-winds 
(GSW, e.g. Cen \& Ostriker, 2006) already at redshifts of $\sim 0.5$, remaining essentially unchanged from then to $z\simeq 0$. 
The first pieces of observational evidence of this violent process of structure formation are starting to be gathered in the densest nodes of this 
filamentary IGM, at and inside the virial radius of massive clusters of galaxies (e.g. Keshet et al. 2012).  
However, diffuse WHIM filaments are too tenuous to be detected through their bremsstrahlung and line emission with current Far-Ultraviolet (FUV)  or X-ray 
instruments (e.g. Yoshikawa et al. 2003). Absorption is more promising. 

The lowest temperature tail of the WHIM distribution, from T$\sim 10^{5}$ up to $\sim 10^{5.5}$ K (where the fraction of OVI peaks), 
has been most likely detected in the FUV, with FUSE and the STIS spectrometer onboard the Hubble Space Telescope ({\em HST}), in OVI and 
narrow Ly$\alpha$ absorbers, though the fraction of this gas that actually belongs to the shock-heated WHIM is highly uncertain (e.g. Danforth \& Shull, 2008). 
More recently, the advent of the larger throughput Cosmic Origins Spectrograph of {\em HST} (COS, hereinafter) has enabled the detection of 
somewhat hotter (T$\sim 10^{5}-10^{5.7}$)  and possibly truly shock-heated WHIM, through the detectability of the strongest (i.e. large column density) 
Broad Lyman $\alpha$ absorbers (BLAs) with thermal Doppler parameter $b_{th} = \sqrt{2kT/m_p} \simeq 40-90$ km s$^{-1}$ 
(Danforth, Stocke \& Shull, 2010)
\footnote{The thermal Doppler parameter, is typically narrower than the measured Doppler parameter $b$, which includes a combination 
of internal turbulence and Hubble-flow broadening. Danforth, Stocke \& Shull, 2010, estimate an average correction factor $(b/b_{th}) \simeq 1.2$.} 
. 
These BLAs are in principle good tracers of the warm (T$\sim 10^{5} - 10^{5.7}$ K) portion of the WHIM, where a significant fraction of the WHIM 
($\sim 30-40$\%, e.g. Cen \& Ostriker, 2006) is supposed to reside. 
However, observing $b_{th} \simeq 40-90$ km s$^{-1}$ BLAs requires data of excellent quality and accurate continuum definition to 
detect the low-contrast features. In addition, measuring HI without an ionization correction
\footnote{The measured HI Doppler parameter gives only an upper limit on the gas temperrature, and in addition, the proper ionization correction 
depends also on the gas density because of the second-order contribution of photo-ionization, and this can only be estimated through multi-ion metal 
measurements, e.g. Nicastro et al. 2002).} 
will always underestimate the actual baryon fraction in the fraction of hot IGM sampled by these tracers. 
Given the temperature of this gas, such an ionization correction can only come from the X-ray detection of the highly 
ionized counterparts to such BLAs. 
Finally, the T$\gs 10^{5.7}$ K portion of the WHIM (the vast majority of the WHIM mass) can only be detected through 
highly ionized metals in the X-rays. 

However, detecting the metal tracers of the bulk of the ``Missing Baryons'' in the ``X-Ray Forest'', has proven to be 
extremely difficult. This is because of the unfortunate combination of (a) the still limited resolving power ($R\sim 400$ at 
the location of the OVII K$\alpha$ transition, at $\lambda = 21.6$ \AA) and the low throughput ($A_{\rm Eff} \sim 15-70$ cm$^2$) of the 
current high-resolution X-ray spectrometers [the XMM-{\em Newton} Reflection Grating Spectrometer (RGS, den Herder, et al. 2001) and the 
{\em Chandra} Low Energy Transmission Grating (LETG: Brinkman et al. 2000)]; 
(b) the lack of bright ($f_{0.5-2 keV} \ge 10^{-11}$ erg s$^{-1}$ cm$^{-2}$) extragalactic point-like 
targets at sufficiently high redshifts; and, (c) the dramatic steepening in slope ($\Delta\alpha \gs 1.5$) of the predicted number density of metal He-like WHIM 
filaments per unit redshift at ion column densities $\gs 10^{15}$ cm$^{-2}$ (e.g. Cen \& Fang, 1996). 
Current evidence is still limited, and highly controversial. 

The statistically strongest evidence to date comes from the evidence for large amounts of baryonic matter 
in two intervening WHIM filaments along the line of sight to the nearby ($z=0.03$) blazar Mkn~421 (Nicastro et al. 2005a,b). 
However, these results are controversial (Kaastra et al. 2006, Rasmussen et al. 2007, Danforth et al. 2011, Yao et al. 2012; 
but see also Williams et al. 2006, Nicastro, Mathur \& Elvis, 2008, Williams et al. 2010). 

Over the last few years there have been new discoveries reported (Buote et al. 2009: B09; Fang et al. 2010: F10; 
Nicastro et al. 2010, Zappacosta et al. 2010, Zappacosta et al. 2012). 
However, such tentative discoveries are either serendipitous (and so biased toward the strongest systems: Nicastro et al. 2010; 
Zappacosta et al, 2012), or trace extreme galaxy overdensity regions (Zappacosta et al. 2010) or even just galaxy halos (Buote et al. 2009; 
Zappacosta et al. 2010; Williams et al. 2012). In all cases, these sample only the hottest and densest IGM or circumgalactic medium, which is 
not representative of the majority of the WHIM. 

Thus a firm, high-significance detection of the bulk of the missing baryon mass has not been secured yet. 
To do so, the only pursuable observational strategy, with current instrumentation, is to first identify ``true'' gaseous 
signposts (e.g. Mathur, Weinberg \& Chen, 2003) of the T$\sim 10^{5.3} - 10^{5.7}$ WHIM (i.e. BLAs), and then search for their metal counterparts 
in the soft X-rays, along optimally selected sightlines. 
Over the past two years we selected 1ES~1553+113 as ``the best WHIM target in the Universe' (because of its extreme 
soft X-ray brightness, relatively high redshift and the presence of BLA signposts in its {\em HST}-COS spectrum - Danforth et al. 2010), 
and started the X-ray campaign with a pilot 500 ks {\em Chandra} cycle 12 observation of 1ES~1553+113. 
In our observing proposal, we conservatively tuned the exposure of this first pilot observation to test the reliability of the strongest FUV 
systems as gaseous signposts for X-ray WHIM filaments. 
The results are well beyond the most optimistic expectations and confirm the effectiveness of such an observational strategy. 
In this article we report on these results from the pilot {\em Chandra}-LETG spectrum. 

The paper is organized as follow. In \S 2 we describe the target of our LETG observation, and summarize the relevant results from the 
{\em HST}-COS observations reported by Danforth et al. (2010). Section 3 describes the {\em Chandra}-LETG data reduction and analysis, 
while in \S 4, we discuss our findings and present possible absorption system identifications.  
Finally in \S 5, we summarize our conclusions. 
In a forthcoming article (``paper II''), we will present a self-consistent modeling of the spectral features detected in the LETG spectrum 
of 1ES~1553+113 by making use of our WHIM hybrid collisional-ionization plus photoionization models, and so deriving preliminary estimates 
of metallicities of these systems, the ionization-corrected baryon cosmological mass density of the WHIM at T$\sim 10^{5.5}$ K, the number 
density of CV and OVII WHIM absorbers as a function of the ion column density. 

\section{1ES~1553+113 and the WHIM Signposts along its Line of Sight}
Four years ago, we selected 1ES~1553+113 as the optimal target for FUV and X-ray WHIM studies, because of its relatively high estimated 
redshift prior to spectroscopic determinations ($z\gs 0.25$, from the non-detection of the host galaxy in a deep {\em HST} R image: Treves et al. 2007), 
its high Galactic latitude ($b \simeq 44^{\circ}$) and its extreme average 0.1-2.4 keV brightness of 2 mCrab ($4 \times 10^{-11}$ erg s$^{-1}$ cm$^{-2}$), 
coupled with a bright and stable FUV flux of $\nu f_{\nu}(1600\AA) \simeq 3 \times 10^{-11}$ erg s$^{-1}$ cm$^{-2}$. 

1ES~1553+113 was first observed with the {\em HST}-COS on September 22, 2009, for 3 orbits (1 orbit with the G130M 
and 2 with the G160M grating: Danforth et al. 2010). Almost 2 years later, on July 24, 2011, the target was re-observed 
for 2 additional orbits with the G130 M and 4 additional oribits with the G160M. 
The first COS spectrum of 1ES~1553+113 was published in 2010, by Danforth and collaborators. 
Here we summarize the main findings from this 2009 observation, relevant to this work.

The published {\em HST}-COS spectrum of 1ES~1553+113 secured a robust lower limit on the spectroscopic redshift of 1ES~1553+113.  
Narrow HI Ly$\alpha$ (NLA) and OVI absorption is observed at a combined 19.3$\sigma$ significance at $z=0.395$ (Fig. 2 of Danforth et al. 
2010), so establishing a spectroscopic lower limit on the redshift of 1ES~1553+113 of $z\gs 0.4$ (almost twice the non-spectroscopic 
value upon which we based our original selection) and making this target by far the brightest, non-transient, X-ray source  in the $z \ge 0.4$ sky. 
Moreover, Danforth et al. (2010) report on the presence of 5 possible BLAs, at $z=0.04281$, 0.10230, 0.12325, 0.13334 and 0.15234, plus 
a rare and complex triple-HI-metal absorption system at $z=0.1864-0.1899$, spanning a range of 1200 km s$^{-1}$ in velocity (Fig. 2, Table 1, 
in Danforth et al. 2010). 
In our work, we use these systems as initial gaseous signposts for our X-ray search of either WHIM filaments (BLA counterparts), or highly ionized 
galaxy outflows and  hot circumgalactic medium from nearby galaxies (counterparts to the triple-HI-metal system). 
However, since the 2011 COS observation observation almost doubled the S/N of the first published spectrum of 1ES~1553+113, we also 
retrieved and reduced the total final COS spectrum of 1ES~1553+1§13, and present here a re-analysis of it whenever the higher S/N 
data detect FUV lines associated to the proposed X-ray identifications and that had eluded detections in the published spectrum because of its 
lower S/N. 

\section{The {\em Chandra}-LETG spectrum of 1ES~1553+113}
{\em Chandra} observed 1ES~1553+113 between May-June 2011, for 500 ks, with the LETG spectrometer (constant resolution 
$\Delta \lambda = 50$ m\AA, i.e. resolving power R$=400$ at $\lambda = 20$ \AA) dispersed across the High Resolution Camera for Spectroscopy 
(HRC-S, Murray et al. 1985) plates. 
The observation consisted of three visits with exposures of 166.3 ks (2011, May 4-6), 175.4 ks (2011, May 6-9) and 
153.9 ks (2011, June 18-20). 

\subsection{Data Reduction}
The data were reduced with the latest version of the Chandra Interactive Analysis of Observation software (CIAO v. 4.4.1, CALDB v. 4.5.2), 
following the standard processing procedures outlined in the HRC-S-LETG grating analysis thread
\footnote{http://cxc.harvard.edu/ciao/guides/gspec\_hrcsletg.html}
. 

The {\em Chandra} HRC-S detector has virtually no spectral resolution, which prevents the separation of the different spectral orders dispersed 
by the LETG. 
For each observation, we therefore extracted “all-order” negative and positive source and background spectra with the CIAO tool {\em tg\_extract}. 
This produced a total of six LETG spectra, three (one per LETG visit) negative-order and three positive-order source and background spectra, 
extracted at a binning pace of $\Delta\lambda = 12.5$ m\AA, i.e. oversampling the LETG FWHM resolution of 50 m\AA\ by a factor of 4. 

Due to the impossibility of separating HRC-LETG spectral orders, the spectral modeling of these spectra must be performed by folding 
the fitting models with the sum of the convolution products of the redistribution matrices (RMFs) and ancillary response files (ARFs) belonging 
to the first N positive and negative orders, respectively. 
For our LETG observations of 1ES~1553+113, we conservatively decided to build our final “all-order” response matrix by adding up positive and 
negative orders up to N = 10. For each order, we used the CIAO tools {\em mkgarf} and {\em mkgrmf} to build 20 effective area and 
photon-redistribution matrices, which were then read together with the data into the CIAO fitting-package {\em Sherpa}, through the 
commands {\em load\_multi\_arfs} and {\em load\_multi\_rmfs}. 

To maximize the S/N of our spectra (we note that the HRC-S-LETG combination has an effective area of only A$_{\rm Eff} \simeq 15$ cm$^2$ at 
both $\lambda = 22$ and 44 \AA), we first co-added the three negative and three positive-order spectra 
(source and background) from the three observations (CIAO tool {\em add\_grating\_spectra}), to produce 2 LETG spectra (one negative- and one 
positive-order spectrum), with net exposures of 495.6 ks each. 
Effective area matrices of the three observations, for each of the 10 negative and ten positive orders, 
were averaged together, producing 20 averaged effective area matrices (ARFs). 
Finally, we used the CIAO tool {\em add\_grating\_orders} to co-add the two negative and positive order source and background spectra and 
ARFs, to end up with a single LETG spectrum, containing all counts from the ten negative-order and the ten positive-order spectra of the 
three {\em Chandra} observations of 1ES~1553+113. 
The final co-added spectrum, has roughly 250-300 net (i.e. background subtracted) counts per $\Delta\lambda = 50$ m\AA, resolution 
element, over the two interesting spectral bands covering the $z<0.4$ OVII K$\alpha$ ($\lambda = 21.6 - 30$ \AA) and CV K$\alpha$ 
($\lambda = 40.2 - 56$ \AA) ranges at redshifts where WHIM absorption might be found, for a signal-to-noise ratio per resolution element 
(S/N)$\simeq 16$. 

Because of our co-adding procedure, particular care must be taken to account for the wavelength calibration inaccuracy. 
The {\em Chandra} LETG-HRC-S spectrometer suffers a variable intrinsic wavelength uncertainty owing to the non-linearity in the effective
dispersion relation. This non-linearity is due to a spatial imaging non-linearity of the HRC-S detector caused by the combination of (1)
partial charge loss in the 3-tap read-out algorithm of each of the three HRC-S microchannel plates, each of which is divided into 190
taps along the dispersion direction, and (2) a hardware problem that corrupts the data from the position taps under a specific set of
conditions.  The net result is that photon event positions obtained from processing HRC-S telemetry can be offset from their true values
\footnote{See http://cxc.harvard.edu/cal/Letg/Hrc\_disp/degap.html}
. The wavelength errors thus depend on the physical location of dispersed photon events on the HRC-S, and so on both precise nominal
pointing direction and the deliberate Lissajous figure ``dither'' about this point
\footnote{See, e.g. http://cxc.harvard.edu/cal/Letg/Corrlam/}
. 
An updated version of the degap polynomial coefficients based on empirical wavelength corrections from multiple HRC-S/LETG observations 
of Capella has been released since CALDB 3.2.15. 
This update allows the correction of the non-linearities in the HRC-S dispersion relation improving the 90\% statistical significance uncertainties 
across the detector (i.e. $1.6\sigma$) from 22 m\AA\ to 16 m\AA. 
Since we have co-added three different 
observations, and then the resulting positive and negative orders, we should consider a larger uncertainty. 
By propagating four times, in quadrature, the quoted deviations, we obtain 90\% uncertainties of 32 m\AA. 
However, since the updated wavelength corrections have been applied to the rest-frame position of the strongest soft X-ray metal 
electronic transitions and the interpolation of the correction may not be strictly valid to the position of blueshifted and redshifted lines, 
we conservatively adopt a 90\% wavalength uncertainty 40 m\AA, 80\% of the LETG Line Spread Function (LSF) FWHM. 

In the following we report all uncertainties at the 1$\sigma$ (68\%) statistical significance level, with the exception of X-ray wavelength and 
redshifts, for which we use 90\% statistical confidences (i.e. 1.6$\sigma$). 

\subsection{Data Analysis}
We used the fitting package {\em Sherpa}, in CIAO, to fit our {\em Chandra}-LETG spectrum of 1ES~1553+113. 
We ignored data outside the $10 - 60$ \AA\ spectral interval, where the (S/N) was either too low ($\lambda > 60$ \AA\ and 
$\lambda < 5$ \AA) and/or the spectral resolving power too low (at $\lambda < 10$ \AA, R$<200$, implying $\Delta v > 1500$ km s$^{-1}$). 

We first fitted the spectrum with a model consisting of a single power law absorbed by neutral gas. 
This yielded a best-fitting power-law energy spectral index and cold-absorber column density of $\Gamma = 2.66 \pm 0.01$ 
and N$_H = (3.71 \pm 0.05) \times 10^{20}$ cm$^{-2}$ (fully consistent with the Galactic value along the line of sight to 1ES~1553+113: 
N$_H^{Gal} = 3.65 \times 10^{20}$ cm$^{-2}$ (Kalberla et al. 2005). 
The integrated reduced $\chi^2$ was acceptable, $\chi^2_r (dof) = 0.90 (3995)$. However, a visual inspection of the residuals showed 
both broad deviations around the K instrumental edges of OI ($\lambda = 23.3$ \AA) and CI ($\lambda = 43.6$ \AA), the M edges of I and Cs 
($\lambda \simeq 11.5 - 12.4$ \AA) and the HRC-S plate gaps ($\lambda \simeq 48-58$ \AA), as well as narrow 
deficit of counts at specific wavalengths. Figure 1, top panel, shows these residuals smoothed through a convolution with the LETG LSF, 
in re-normalized (after smoothing) standard deviations. 

At first, we excluded the possible absorption lines (which do not affect the continuum fitting over the integrated 10-60 \AA\ band) and 
focused on curing the residual calibration uncertainties around the positions of the instrumental I, Cs, O and C edges and the HRC-S plate 
gaps, with the aim of reaching an optimal model of the continuum that would allow us to properly assess the statistical significance of true physical 
spectral features possibly present in the spectrum. 
We then added to the continuum model, (a) an edge with negative optical depth, to cure the sharp edge-like emission feature present in the 
residuals at $\lambda \simeq 43.8$ \AA\ and due to either an over-estimated modeling of the CI instrumental edge in the LETG effective 
area, or a non-accurate modeling of the CI absorption edge in the {\em wabs} model of XSPEC, or both (best-fit $\tau = -0.04$); 
(b) three broad emission Gaussians at $\lambda = 41.61$ \AA, $\lambda = 50.10$ \AA\ and $\lambda = 56.39$ \AA\ (best-fitting $\sigma = 0.8$ 
\AA\ for the first two and $\sigma = 0.2$ \AA\ for the third) to model residuals broad wiggles at the wavelength of the CI edge, and the 
dithering-shaped HRC-S plate gaps
\footnote{During every pointing the {\em Chandra} satellite dithers around the aim-point sky coordinates producing a  Lissajous-figure in detector 
coordinates. This implies that plate-gaps are not blind to photons at all times during the observation, and their dispersion position (for dispersed 
spectra) can be reconstructed a posteriori by applying the aspect-solution. However, the exact shape of the effective area at the plate-gap 
wavelengths depends on the accuracy of the aspect-reconstruction and the zeroth-order centroid.}
; (c) Five {\em notch} XSPEC component models: 4 negative (at $\lambda = 11.48$, $\lambda = 43.99$, 48.34 and 52.14 \AA, with total widths 
of 0.6, 0.4, 0.2 and 1.1 A, respectively), to model residual narrower (but still well resolved) deviations at the I and Cs M edges and the C K edge 
and plate-gaps wavlengths Fig. 1, top panel), and one positive with central $\lambda = 20.14$ \AA\ and total width of 5.5 \AA, to model a 
systematic excess over the 18-23 \AA\ spectral region, just leftward of the instrumental O K edge.

We re-fitted the data, and obtained an improvement of $\Delta \chi^2 = 107$, for 20 additional degrees of freedom, with almost unchanged 
best-fitting parameters: $\Gamma = 2.62 \pm 0.01$ and N$_H = (3.60 \pm 0.05) \times 10^{20}$ cm$^{-2}$. We measure a (0.1 - 2) keV flux 
of F$_{0.1-2} = (2.45 \pm 0.03) \times 10^{-11}$ erg s$^{-1}$ cm$^{-2}$ (1.2 mCrab). 
The LSF-smoothed residuals now appear flat over the entire 10-60 \AA\ band (Fig. 1, bottom panel). 

%-----------------------------Figure Start------------------------------
\begin{figure}[h]
\begin{center}
\hbox{
% un-comment the following line to include your fig1a.ps postscript file
\hspace{1.0cm}
\psfig{figure=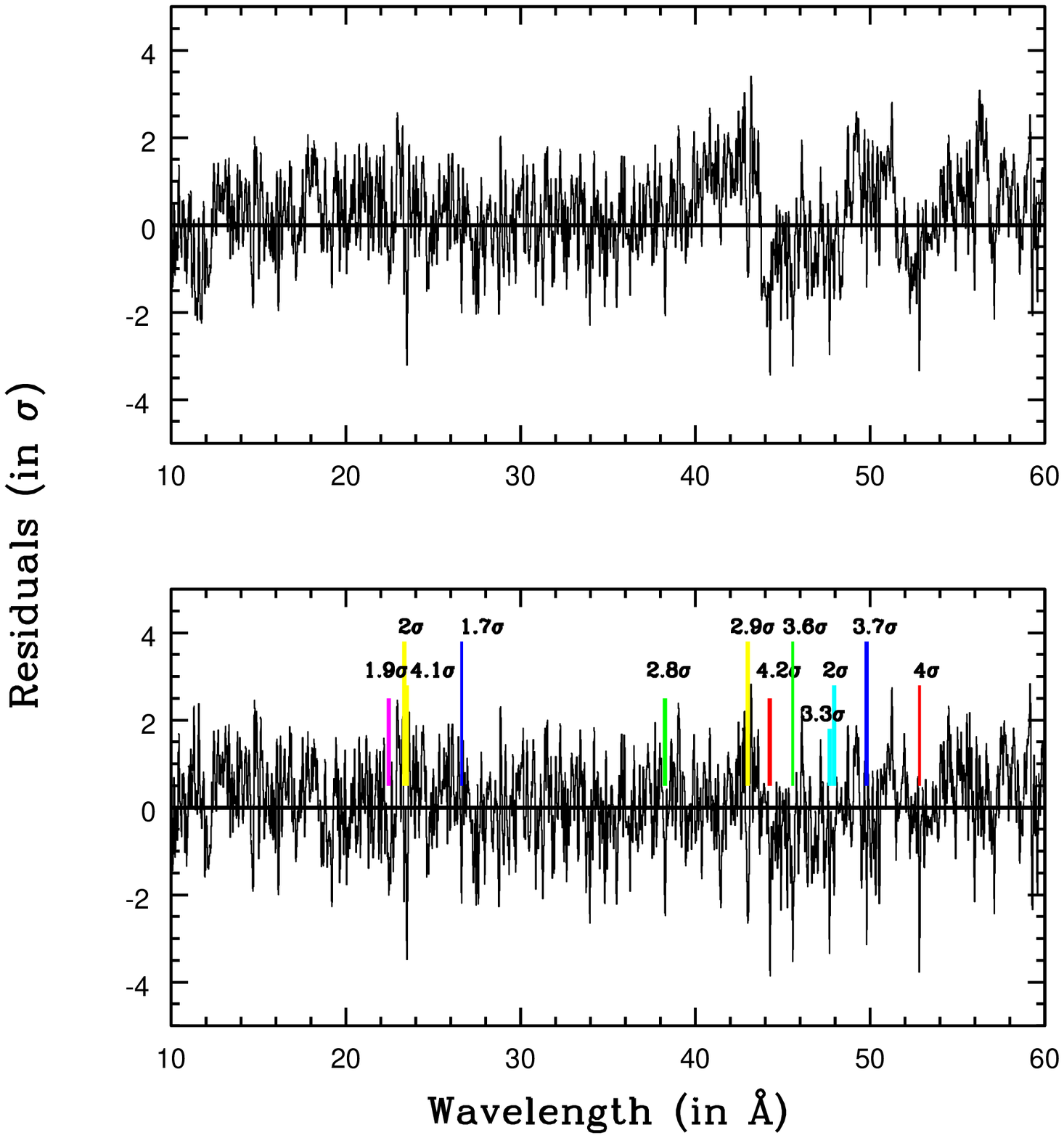,height=15.0cm,width=15.0cm,angle=0}
% un-comment the following line to include your fig1b.ps postscript file
%\psfig{figure=fig1b.ps,height=7.0cm,width=7.0cm,angle=-90}
}
\end{center}
\vspace{-1.5cm} 
\caption{\footnotesize LSF-smoothed residuals (in re-normalized, after smoothing, standard deviations) of the 10-60 \AA\ 500 ks LETG 
spectrum of 1ES~1553+113, to: (a) a model consisting of a power-law absorbed by a neutral absorber (top panel), and (b) the same continuum model 
with the addition of 7 components (an edge with negative optical depth, two broad Gaussians, three negative and one positive {\em notches}, to cure 
the residual systematic calibration uncertainties visible in the top panel and due to the instrumental OI and CI edges and the dithering-shaped 
HRC-S-LETG effective area at the wavelengths of the HRC-S plate gaps (bottom panel). 
In the bottom panel, the positions of the 12 possible absorption lines are marked with color-coded segments (\S 4) and labeled with their 
single-line statistical significances.}
\label{fig1}
\end{figure}
%-----------------------------Figure End--------------------------------

\subsubsection{Unresolved Absorption Lines: Detectability of Carbon vs Oxygen}
A visual inspection of the LSF-smoothed residuals in Fig. 1 (bottom panel), shows clear narrow (i.e. unresolved: $\Delta \lambda \ls 0.05$ 
\AA) deficit of counts involving 3-6 consecutive bins, especially in the spectral region where redshifted C absorption is expected. 
This is expected: indeed, the sensitivity of dispersed spectrometers with constant $\Delta\lambda$ and S/N per resolution element, over a 
given band-pass, to a given ion column density, increases with wavelengths. 
This is because of the concurrence of two independent facts. 
First, the absorption line equivalent width (EW) sensitivity of a dispersed spectrum with given constant $\Delta\lambda$ and S/N, down to a statistical 
significance N$\sigma$, can be written: 
\begin{equation}
EW_{thresh} = N_{\sigma} [\Delta\lambda / (S/N)]. 
\label{eq1}
\end{equation}
Second, the absorption line EW is related to the column density of the ion imprinting the line in the spectrum through the rest-frame position 
of the line and the redshift of the absorber. For unsaturated lines we can write: 
\begin{equation}
EW(X^i) \simeq 8.9 \times 10^{-21} \times N_{X^i} \xi_{X^i} \lambda_{rf}^2 (1+z), 
\label{eq2}
\end{equation} 
with $EW$ in \AA\ and where $N_{X^i}$ is the column density of the ion $i$ of the element $X$ in cm$^{-2}$, $\xi_{X^i}$ is the oscillator strength of the 
given transition of the ion $X^i$, $\lambda_{rf}$ the rest-frame wavelength (in \AA) of the given transition of the ion $X^i$, and $z$ the redshift of the 
absorber. 
Therefore, from equations \ref{eq1} and \ref{eq2}: 
\begin{equation}
N_{X^i}^{Thresh} \simeq 1.1 \times 10^{20} \times N_{\sigma} \xi_{X^i}^{-1} \lambda_{rf}^{-2} (1+z)^{-1} \Delta\lambda (S/N)^{-1}. 
\label{eq3}
\end{equation}
Let us now consider, the LETG spectrometer and, for example, the strongest K$\alpha$ transitions of the two stable and most abundant ions of carbon 
and oxygen at T$\simeq 10^6$ K, i.e. CV and OVII. The rest frame wavelengths of these two transitions are in a 1:2 ratio, and fall in regions of the 
spectrometer with similar effective area (i.e. similar S/N in the same spectrum). Therefore, from Eq. \ref{eq3}, one gets 
$$[N_{CV}^{Thresh} / N_{OVII}^{Thresh}] \simeq [\lambda(CV K\alpha) / \lambda(OVII K\alpha) ]^{-2} \simeq 1/4.$$ 
For a broad range of WHIM temperatures, logT$\simeq 5.5 - 6$, the fraction of CV and OVII are similar and between 0.7-1. At $5 <$ logT $< 5.5$ 
CV largely dominates over OVII. Moreover, in gas with solar-like composition, oxygen is roughly twice more abundant than carbon. So, for $5 <$ 
logT $< 6$ the minimum detectable CV column density is at least twice smaller than the minimum detectable OVII column density.

\subsubsection{Automated Identification of Absorption-Line Candidates}
To identify candidate unresolved absorption lines we wrote a routine that scans the LSF-smoothed residuals (Fig. 1, bottom panel) over 
$\Delta\lambda = 125$ m\AA\ wide contiguous bins (2.5 LETG resolution elements), to search for negative deficits of counts involving 
between 3-7 consecutive bins (i.e. 0.75-1.75 LETG LSF FWHM at our oversampled resolution), and line-shaped, i.e. symmetrically distributed 
around a central negative peak (at least one left and one right bin with values higher than the peak).  
We do not consider resolved lines here (i.e. consecutive deficits of counts spreading over more than $\simeq 7$ bins), since we 
assume that any ISM or IGM absorption lines must have intrinisc FWHM $\ll 800$ km s$^{-1}$. 
By definition, LSF-smoothing of the residuals tends to smooth flickering in adjacent sub-resolution bins. Thus, on the one hand it highlights 
real absorption (or emission) lines, but on the other hand it also increases the number of false absorption (or emission) line-like features with 
typical LSF width (see, e.g., Fig. 1 bottom panel). 
We can estimate the expected number of LSF-shaped emission or absorption features exceeding a bin-by-bin integrated (in quadrature) 
significance of 3$\sigma$, by using a simple Gaussian argument. At our 4$\times$ LSF over-sampled binning, we have a total of 
(60-10)/0.0125 = 4000 bins. Therefore the Gaussian-predicted number of positive or negative LSF-shaped fluctuations at $\ge 3\sigma$ is 
(1 - 0.997)$\times 4000$ = 12. So, on average, we expect a total of 6 emission and 6 absorption LSF-shaped fluctuations.
Indeed, the blind run of the scanning routine over the 10-60 \AA\ LSF-smoothed residuals (over-sampled at a resolution of 0.25$\times$ the 
LETG LSF FWHM), finds exactly 6 LSF-shaped emission features exceeding the bin-by-bin integrated (in quadrature) statistical significance of 
$3\sigma$, according to expectations given the blazar nature of the target and the consequent absence of intrinsic emission in the optical 
or infrared spectra of 1ES~1553+113. 
On the contrary, 50 of such features are found in absorption, clearly suggesting that real absorption lines are imprinted on the  
10-60 \AA\ LETG continuum spectrum of 1ES~1553+113. 
Given the way the scanning routine works, moving on contiguous 2.5 LETG-LSF wide bins, some of the 50 absorption line candidates found 
by the routine are simply the same strong feature, detected twice over adjacent scanned bins. This leaves about 40 absorption-line 
candidates, still many more than predicted by Gaussian statistics and found in emission. 

To confirm the real absorption-line nature of these candidates (at least down to the sensitivity of the current LETG spectrum), we therefore 
used unresolved (line FWHM frozen at 5 m\AA) absorption Gaussians to model the non-smoothed data. 
We added these Gaussians one by one and refitted the data, by allowing the line position to vary within 0.5 \AA\ around the scan-routine 
determined position of the centroid. 
This procedure yielded the results summarized in Table 1, columns 1 and 2. Only 12 of the possible $\sim 40$ absorption candidates were actually 
confirmed by the fitting procedure at $> 90$\% ($1.6\sigma$) statistical significance. 
The single-line (prior to secure identification) statistical significances of 11 of these lines (computed as the ratio between 
the best-fitting line normalization and its 1$\sigma$ uncertainty) are in the range  $(2.2 - 4.1)\sigma$: three lines at $> 4\sigma$, 
three at $3.6 \le \sigma < 4$, five at $2.2 \le \sigma \le 2.8$. (Table 1, column 2 and Fig. 1 bottom panel).
An additional line-like feature is only hinted by the data and can be fitted with a negative Gaussian with single-line statistical significance of 
1.7$\sigma$. 
In the following we consider the six lines with prior-to-identification statistical significance $\ge 3.6\sigma$ (0.04 and 0.08 expected 
by chance up to $z=0.4$ in the oxygen and carbon bands, respectively; see \S 4.2) as real absorption lines, which need to be identified. 
The remaining six absorption features are considered as tentative detections that need confirmation, and we consider them here only 
in association with either higher significance LETG absorbers or COS signposts, or both. 
In Figure 2 we show the best fitting models of the 12 absorption lines superposed on the non-LSF-smoothed residuals (in standard 
deviations) of the data to the best-fitting continuum model. Best fitting Gaussians models are color-grouped into 6 different systems, 
according to their possible identifications (\S 4.1). 

%%%%%%%%%%
\begin{table}
\footnotesize
\begin{center}
\caption{\bf Absorption Lines in the LETG (and COS) spectrum(a) of 1ES~1553+113}
\vspace{0.4truecm}
\begin{tabular}{|cccc|c||}
\hline
$^a$Wavelength & $^b$Significance & Redshift & Id & COS HI, Metal I \\
in \AA\ & in $\sigma$ & & & \\ 
\hline
\multicolumn{5}{|c|} {Galactic Absorbers} \\
\hline 
$23.34 \pm 0.04$ & 2.5 & $\pm 0.002$ & OI Molecular & N/A \\
$23.49 \pm 0.04$ & 4.1 & $-0.001 \pm 0.002$ & OI K$\alpha$ Atomic & N/A \\
$43.01 \pm 0.04$ & 2.8 & $\pm 0.001$ & CII K$\alpha$ Atomic & N/A \\
\hline
\multicolumn{5}{|c|} {Intervening Absorbers} \\
\hline 
\multicolumn{5}{|c|} {High-Confidence X-ray System Detections and COS Associations} \\
\hline 
$44.27 \pm 0.04$ & 4.1 & $0.312 \pm 0.001$ & CVI K$\alpha$ & BLA-OVI $z_{FUV} = 0.3113$  \\
$52.82 \pm 0.04$ & 4.1 & $0.312 \pm 0.001$ & CV K$\alpha$ & BLA-OVI $z_{FUV} = 0.3113$ \\
\hline
$49.80 \pm 0.04$ & 3.9 & $0.237 \pm 0.001$ & CV K$\alpha$ & BLA $z_{FUV} = 0.23666, 0.23734$  \\
$[$ $26.65 \pm 0.04$ & 1.7 & $0.234 \pm 0.002$ & OVII K$\alpha$ & BLA $z_{FUV} = 0.23666, 0.23734$ $]^c$ \\
\hline
$45.58 \pm 0.04$ & 3.8 & $0.132 \pm 0.001$ & CV K$\alpha$ & BLA$^d$ $z_{FUV} = 0.1333$ \\
$38.25 \pm 0.04$ & 2.7 & $0.134 \pm 0.001$ & CVI K$\alpha$ & BLA$^d$ $z_{FUV} = 0.1333$ \\
\hline
\multicolumn{5}{|c|} {Tentative X-ray Detections and COS Associations} \\ 
\hline 
$47.68 \pm 0.04$ & 3.6 & $0.184 \pm 0.001$ & CV K$\alpha$ & Triple-HI: $z_{FUV}=0.186-190$ \\
$47.94 \pm 0.04$ & 2.2 & $0.191 \pm 0.001$ & CV K$\alpha$ & BLA-OVI $z_{FUV}=0.18989$ \\
$[$ $26.65 \pm 0.04$ & 1.7 & $0.191 \pm 0.002$ & OV K$\alpha$ & BLA-OVI $z_{FUV} = 0.18989$ $]^c$ \\
\hline
$22.48 \pm 0.04$ & 2.3 & $0.041 \pm 0.002$ & OVII K$\alpha$ & BLA $z_{FUV}=0.0428$ \\
\hline
\end{tabular}
\end{center}
$^a$ We use 40 m\AA\ as 1$\sigma$ systematic uncertainities for the HRC-LETG co-added negative and positive order wavelength scale. 
This is (4/5) of the LETG Line Spread Function FWHM, and it is due to the non-linearity in the HRC-LETG effective dispersion relation. 

$^b$ Single-line statistical significances, obtained by ratioing the best-fitting line normalization and its 1$\sigma$ error. 

$^c$ This line can be identified with either an OVII K$\alpha$ transition at $z_X = 0.234 \pm 0.002$, close but only 
3$\sigma$ consistent with the CV-BLA system at $z_X = 0.237 \pm 0.001$ (\S 4.1.3), or with an OV K$\alpha$ transition associated with the 
CV-BLA-OVI system at $z_X = 0.191 \pm 0.001$ (\S 4.1.5). 
 
$^d$ Given the relatively low-temperature derived from the HI Doppler parameter, the column density of this HI component is 
probably too low to produce all the CV and CVI detected in the LETG, even assuming Solar abundances. (see \S 4.1.3).
\end{table}

%\end{sidewaystable}
\normalsize
%%%%%%%%%%%

%-----------------------------Figure Start------------------------------
\begin{figure}[h]
\begin{center}
\hbox{
% un-comment the following line to include your fig1a.ps postscript file
\hspace{1.0cm}
\psfig{figure=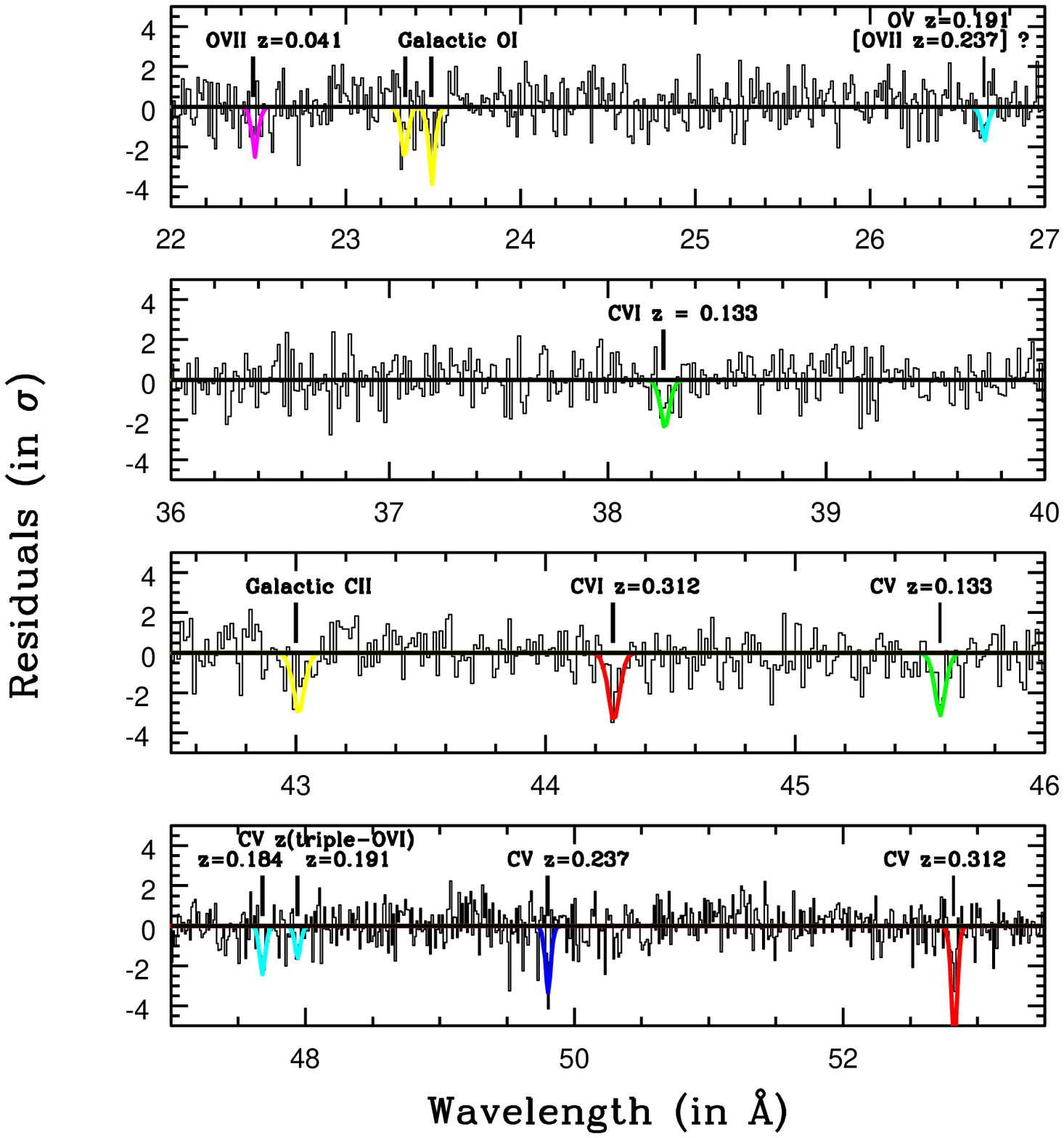,height=15.0cm,width=15.0cm,angle=0}
% un-comment the following line to include your fig1b.ps postscript file
%\psfig{figure=fig1b.ps,height=7.0cm,width=7.0cm,angle=-90}
}
\end{center}
\vspace{-1.5cm} 
\caption{\footnotesize Non LSF-smoothed residuals (in $\sigma$) of four portions of the 500 ks LETG spectrum of 1ES~1553+113, where Galactic 
OI and CII, as well as intervening OVII and CV, are tentatively detected, at single-line statistical significances in the $2.2-4.1\sigma$ range. 
Best-fitting curves of different colors identify different systems: (1) Galactic OI and CII (yellow), (2) OVII at $z_X=0.041$ (magenta), (3) CV 
(and myabe OVII) at $z_X =0.237$ (blue), (4)  CVI and CV at $z_X=0.312$ (red), (5) CV and CVI at $\langle z_X \rangle=0.133$ (green), and 
(6) CV at the COS triple-HI-metal redshifts (cyan).}
\label{fig2}
\end{figure}
%-----------------------------Figure End--------------------------------

\section{Discussion}
The LETG spectrum of 1ES~1553+113, reveals the presence of unresolved line-like absorption in both the $z<0.4$ 
oxygen K$\alpha$ and carbon K$\alpha$ regions. These absorption lines could be the imprints of either Galactic ISM and/or CGM, or 
of warm-hot intervening IGM, or both. 
To identify these systems, we consider all the strongest ground-state metal transitions (K$\alpha$) from all ions of the two most abundant 
elements in gas with solar-like composition (O and C). We try to match the possible identifications, with the available redshift priors: 
(a) $z=0$, rest-frame absorption by the ISM of our Galaxy, or CGM (the LETG resolution does not allow us to kinematically distinguish between 
the two; e.g. Gupta et al. 2012 and references therein), and (b) the redshifts of the possible BLAs (or NLAs)  and/or OVI absorbers 
identified in the COS spectrum of 1ES~1553+113 (Danforth et al. 2010). 

\subsection{Absorption System Identifications}
The longest wavelength HI Ly$\alpha$ securely identified in the COS spectrum of 1ES~1553+113, is at $\lambda \simeq 1695$ \AA, 
$z=0.395$. At longer wavelengths, and up to the COS G160M band-limit ($\lambda \sim 1800$ \AA, or $z\sim 0.48$ in HI Ly$\alpha$), 
the absorption line density drops dramatically, indicating $z\simeq 0.4$ as a likely redshift for the target (Danforth et al. 2010). 
We therefore assume $z=0.4$ as the redshift of 1ES~1553+113. 

The 12 absorption features identified as possible lines in the LETG spectrum of 1ES~1553+113, have wavelengths between 
$\lambda = 22.48 - 26.65$ \AA\ (four lines in the O-region of the spectrum) and $\lambda = 38.25 - 52.82$ \AA\  (8 lines 
in the C-region of the spectrum). 
The shortest wavelength K$\alpha$ transition of O is that of its H-like ion (OVIII), at $\lambda = 18.97$ \AA\ (Verner, Verner \& Ferland, 1996), 
while the longest wavelength K$\alpha$ transition of O is the inner-shell transition of its neutral ion (OI), 
at $\lambda = 23.52$ \AA\ (E. Behar, private communication)
\footnote{For most of the K$\alpha$ inner-shell transitions of C and O (with the exception of OI and CII, for which empirical observational 
  values are available, e.g. Takei, Fujimoto \& Mitsuda, 2002, Shulz et al. 2003, Juett et al. 2004) we use wavelengths and oscillator strengths 
computed through the HULLAC code (Bar-Shalom, Klapisch \& Oreg, 2001).  
The uncertainties on the rest-frame positions of these transitions are of the order of  0.1-1\% (i.e. $\Delta \lambda \sim 20-200$ m\AA\ (i.e. 
Behar \& Netzer, 2002). However, several of these positions have been bench-marked with either laboratory or astrophysical experiments (e.g. 
Kaastra et al. 2011, and reference therein; Shulz et al. 2003, and references therein; Juett et al. 2004)., and have resulted to be accurate to 
$\Delta \lambda \sim 2-10$ m\AA.}
. Similarly, for C, we have $\lambda(CVI_{L\alpha}) = 33.74$ \AA\ (Verner, Verner \& Ferland, 1996), and $\lambda(CI_{K\alpha}) = 43.37$ and 
43.27 \AA\ (an LETG-resolved doublet; E. Behar, private communication), as the shortest and longest wavelength transitions, respectively. 
Intermediate ionization ions have their K$\alpha$ transition wavelengths between these extremes . 
So, the four LSF-shaped features in the 22-27 \AA\ range (Figure 1, top panel), if true absorption lines, can only be identified with OVIII in 
the ranges $0.185 < z < 0.4$ or with OI-OVII at $0 \ls z<0.4$, with the exact lower limit depending on the ion considered. 
Analogously, the eight line-like features at $\lambda \ge 38.25$ \AA\ can only be identified with CVI in the range $0.134 < z < 0.4$ or CI-CV 
at $0 \ls z < 0.4$, while the feature at $\lambda 38.25$ \AA, can only be due to redshifted CVI.  

We have checked these ions one by one, and propose possible identifications based on HI and/or OVI COS associations. 
We only check here for possible X-ray C and/or O counterparts to the FUV signposts. These are by far the two most abundant metals  
in gas with solar-like composition (Asplund et al. 2009), with C being only a factor of $\sim 2$ less abundant than O, and the third most 
abundant metals (N and Ne) being $\sim 4\times$ less abundant than C. 
Similarly, when in doubt between two or more identifications involving different ions of C and/or O, we always favor the He-like metal 
identification, i.e. CV and/or OVII.  These are the He-like ions of C and O, respectively, and as such are highly stable and by far the most 
abundant ions of their respective elements over a broad range of temperatures. The oscillator strengths of 
their K$\alpha$ transitions are also the strongest of all the other K$\alpha$ transitions from lower- or higher-ionization ions. 

\subsubsection{$z \simeq 0$ Galactic Absorption} 
The three lines at $\lambda = 23.34$ \AA, $\lambda=23.49$ \AA\ and $\lambda = 43.01$ \AA, are consistent with $z \simeq 0$ absorption 
by neutral (or only moderately ionized) species of C and O. 
In particular, we clearly (4.1$\sigma$) detect atomic OI at $\lambda = 23.49 \pm 0.04$ \AA\ and also a marginal (2.5$\sigma$) 
excess of molecular OI (possibly ISM oxygen in coumpound forms of oxide dust grains, e.g. Takei, Fujimoto \& Mitsuda, 2002) at 
$\lambda = 23.34 \pm 0.04$ \AA, with respect to the instrumental molecular OI. 
Atomic OI is not an instrumental feature (e.g. Takei, Fujimoto \& Mistuda 2002). 
Finally we marginally (2.8$\sigma$) detect the signature of CII (whose K$\alpha$ transition is twice as strong as the single transitions 
of the resolved and undetected CI doublet) at $\lambda = 43.01 \pm 0.04$ \AA
\footnote{Not only the CII K$\alpha$ transitions is intrinsically stronger than the CI K$\alpha$, but also CII is expected to be much more abundant than 
CI in the Galaxy, because neutral carbon is easily photoionized by ambient UV radiation (the C I ionization edge is at 11.260 eV or 1101 \AA).} 
. We identify these three transitions as due to the cold ISM of our Galaxy (Table 1, Columns 3-4; Fig. 1 and 2, yellow lines), and we therefore 
consider their single-line statistical significance as their true (i.e. after a secure identifications) statistical significance. 

We also note that we do not detect any OVII K$\alpha$ $z\simeq 0$ absorber along the line of sight to 1ES~1553+113, down to the 
3$\sigma$ sensitivity limit of our Chandra LETG spectrum, in this spectral region: EW(OVII($z=0$))$ < 12$ m\AA. 

Below we present the possible intervening system identifications in order of  statistical significances. 
 
\subsubsection{The $z_X = 0.312 \pm 0.001$ CV-CVI-BLA-OVI WHIM System} 
The line at $\lambda =44.27$ \AA, is that detected at the highest statistical significance ($4.1\sigma$) in the LETG spectrum (together with 
the line at $\lambda = 52.82$ \AA\ and the Galactic OI atomic line). 
The CI, CII and CIV K$\alpha$ transitions, at this wavelength, would be redshifted at $z_X = 0.021$, 0.031 and 0.070. At none of these 
redshifts is there a metal or HI association in COS. 
At the redshift of a putative CIII, instead, $z_X = 0.051$, there is a moderately strong NLA. However, no imprint of a CIII$\lambda 977$ 
absorber is seen in the FUSE spectrum of 1ES~1553+113 (though its presence cannot be ruled out due to the presence of HI Ly$\beta$ and 
OI absorption/airglow in FUSE), and no other metal lines are present in the FUSE or COS spectra at this redshift. 
CV and CVI identifications are more promising. 

The CV K$\alpha$ transition would be redshifted at $z_X = 0.100 \pm 0.001$. At a close (but $> 3\sigma$ inconsistent) redshift 
($z_{FUV} = 0.10230$) there is one of the strongest HI Ly$\alpha$ lines in COS, which is also seen in Ly$\beta$ and is consistent with being 
thermally broadened (BLA) at T$\simeq 10^5$ K. 
Moreover, at $\lambda = 45.58$ \AA\ there is another line in the LETG spectrum (3.8$\sigma$), which, if identified with CIV, would fall at 
$z_X = 0.100 \pm 0.001$. The possible temperature of the HI absorber at $z_{FUV} = 0.10230$ is consistent with the presence of CIV and CV, 
however no CIV doublet ($\lambda = 1548.195, 1550.77$) is seen in the COS spectrum at $z=0.10230$ or $z=0.100$. 
We conclude (both for the lack of CIV in COS and the redshift inconsistency with the BLA) that the $z_X = 0.100$ CV/CIV identification with a 
thick and moderately cool WHIM filament traced by the COS BLA at $z_{FUV} = 0.10230$, is unlikely. 

A more likely identification is with a CVI K$\alpha$ transition at $\lambda = 44.27$ \AA, which would have  a redshift of $z_X = 0.312$. 
At $\lambda = 52.82$ \AA\ there is the other highest statistical significance line of the LETG spectrum (4.1$\sigma$), which, if identified with 
CV K$\alpha$, would fall at $z_X = 0.312 \pm 0.001$, exactly the redshift of the CVI identification. 
No HI or OVI identification is reported at a consistent redshift in the published COS spectrum. However, with additional data, 
the currently available COS spectrum of 1ES~1553+113 has a S/N a factor $\sim 1.5-2\times$ higher than the published spectrum. 
We therefore downloaded from the {\em HST}-MAST archive all available {\em HST}-COS spectra of 1ES~1553+113 taken with the G130M and G160M 
gratings, and re-analyzed the two portions of the spectrum at $1350-1365$ \AA\ and $1590-1600$ \AA, regions where 
the OVI$(1s^22s\rightarrow 1s^22p)$ doublet (i.e. OVI$_1(\lambda 1031.9)$, oscillator strength $f=0.133$, and OVI$_2(\lambda 1037.6)$, $f=0.066$) 
and the HI Ly$\alpha$ at $z\sim 0.308-0.316$ are supposed to lie. 

We identify a strong HI BLA at $z_{FUV} = 0.31124\pm 0.00003$ (8.1$\sigma$; Fig. 3 top panel), with the following best-fitting parameters: 
$\lambda = 1594.03 \pm 0.03$ \AA, $b = 60 \pm 6$ km s$^{-1}$ and EW$= 138 \pm 17$ m\AA. 
The strongest line of the OVI doublet (OVI$_1$) is also detected ($3.9\sigma$; Fig 3, central panel), with best fitting parameters:  
$\lambda = 1353.13 \pm 0.04$ \AA\ (i.e. $z_{OVI} = 0.31130 \pm 0.0004$), $b = 32 \pm 7$ km s$^{-1}$, and EW$= 27.5 \pm 7.0$ m\AA. 
Finally, the weakest line of the OVI doublet (OVI$_2$) is not detected in the data (Fig. 3, bottom panel), but the COS 3$\sigma$ EW upper limit 
at the wavelengths of this line, EW$ < 21$ m\AA, is consistent with the maximum possible EW expected for the OVI$_2$ (given the 
measured EW(OVI$_1$) and the fact that this line is not blank). 
From the observed HI and OVI Doppler parameters, we can estimate the relative contribution of thermal and turbulence broadening. 
In general, if $b_{HI}^{Obs}$ and $b_X^{Obs}$ are the observed Dopller parameters of two associated lines of HI and a metal $X$ with atomic 
weight $A$, then the HI thermal parameter is given by: 
\begin{equation} 
(b_{HI}^{therm})^2 = [A / (A - 1)] \times [(b_{HI}^{Obs})^2 - (b_X^{Obs})^2]. 
\end{equation}
So, in this case we have: $b_{HI}^{therm} = 52 \pm 7$ km s$^{-1}$ (implying logT$=5.2 \pm 0.1$) and $b_{turb} = 30 \pm 14$ km s$^{-1}$, with a 
thermal contribution (in quadrature) of 75\% and a $b_{HI}^{Obs}/b_{HI}^{therm} = 1.15$ correction factor, similar to the average $b_{HI}^{Obs}/b_{HI}^{therm} = 
1.2$ value found by Danforth, Stocke \& Shull (2010). 

We conclude that the two highest significance intervening LETG lines are CV and CVI absorbers at $z_X = 0.312$ (Table 1; Fig. 1 red lines), 
possibly imprinted by an intervening WHIM filament, as further suggested by the a-posteriori discovery of an associated strong BLA and OVI 
counterpart in the COS spectrum, with the right temperature (i.e., line Doppler parameters) and strength. 
This identification is further strenghtened by the a-posteriori discovery of two additional associated X-ray absorption lines. 
Figure 4 shows 5 portions of the LETG spectrum of 1ES~1553+113, centered around the CV K$\alpha$ (top panel), CVI K$\alpha$ (second panel), 
CV K$\beta$ (third panel), OV K$\alpha$, and OVII K$\alpha$ transitions at $z=0.312$. 
Clearly, both the OV K$\alpha$ and the CV K$\beta$ lines are hinted in the data, and were not picked up by our scanning routine because of their bin-by-bin 
integrated (in quadrature) significance, which is slightly lower ($2.7\sigma$ for both lines) than the $3\sigma$ threshold that we imposed in our search 
(\S 3.2). The single-line statistical significance of these two additional lines are 1.9$\sigma$ and 1.6$\sigma$, respectively. 
The OVII K$\alpha$ transition, instead, is undetected (Fig. 4, bottom panel) down to the 3$\sigma$ sensitivity threshold of EW$_{OVII}(z=0.312) =20$ 
m\AA, confirming the relatively low temperature for this system. 

%-----------------------------Figure Start------------------------------
\begin{figure}[h]
\begin{center}
\hbox{
% un-comment the following line to include your fig1a.ps postscript file
\hspace{1.0cm}
\psfig{figure=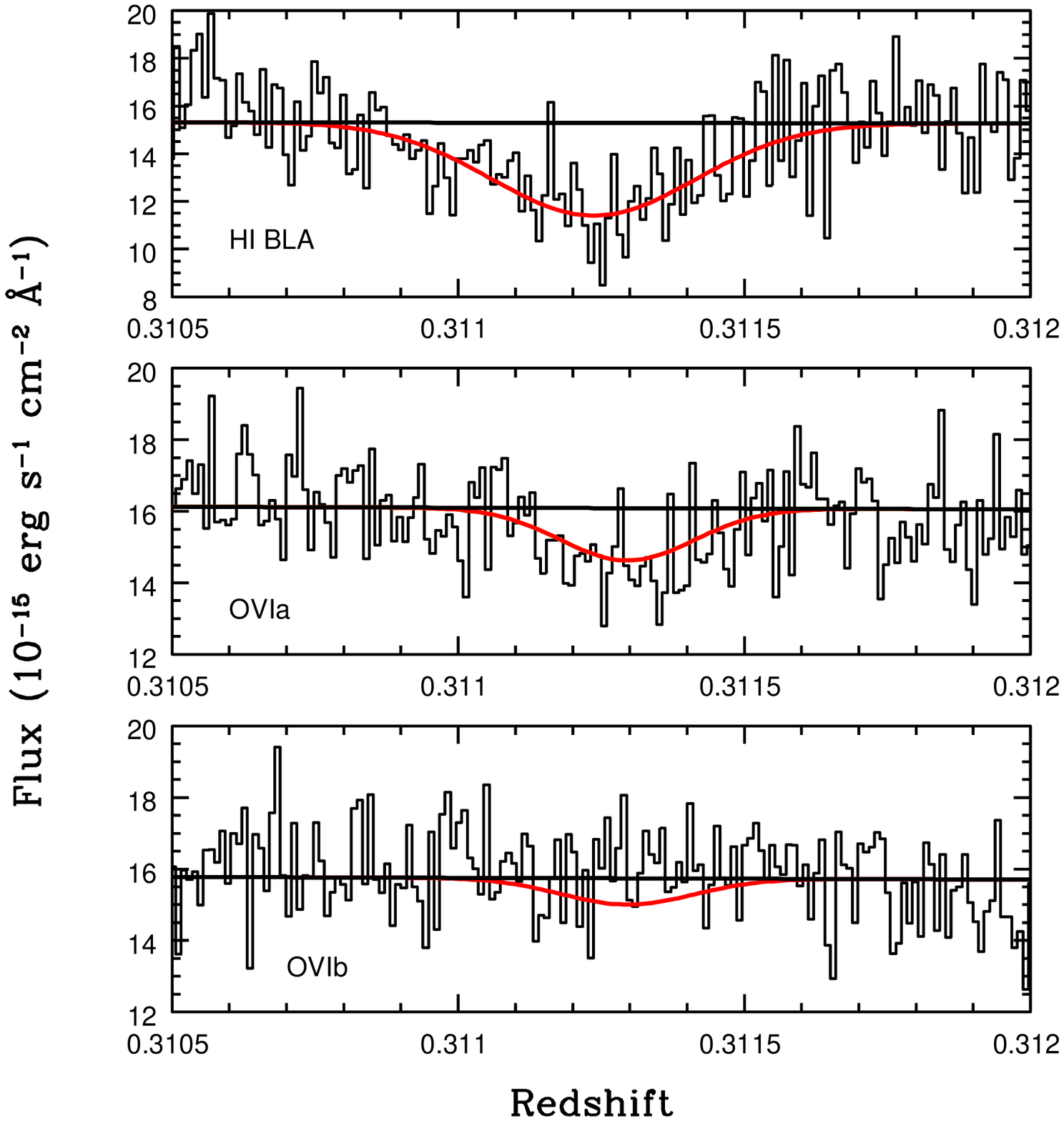,height=15.0cm,width=15.0cm,angle=0}
% un-comment the following line to include your fig1b.ps postscript file
%\psfig{figure=fig1b.ps,height=7.0cm,width=7.0cm,angle=-90}
}
\end{center}
\vspace{-1.5cm} 
\caption{\footnotesize Three portions of the COS spectrum of 1ES~1553+113, centered around the HI Ly$\alpha$ 
(top panel), OVI$_1$ (central panel), and OVI$_2$ (bottom panel), at $z_{FUV} = 0.31124 \pm 0.00003$. 
In all panels, red curves are the best-fitting Gaussians for the HI Ly$\alpha$ (top panel), OVI$_1$ (central panel) and OVI$_2$ (bottom panel). 
COS data are binned at about 1/8 the COS resolution, i.e. $\Delta \lambda \simeq 9$ m\AA.}
\label{fig3}
\end{figure}
%-----------------------------Figure End--------------------------------

%-----------------------------Figure Start------------------------------
\begin{figure}[h]
\begin{center}
\hbox{
% un-comment the following line to include your fig1a.ps postscript file
\hspace{1.0cm}
\psfig{figure=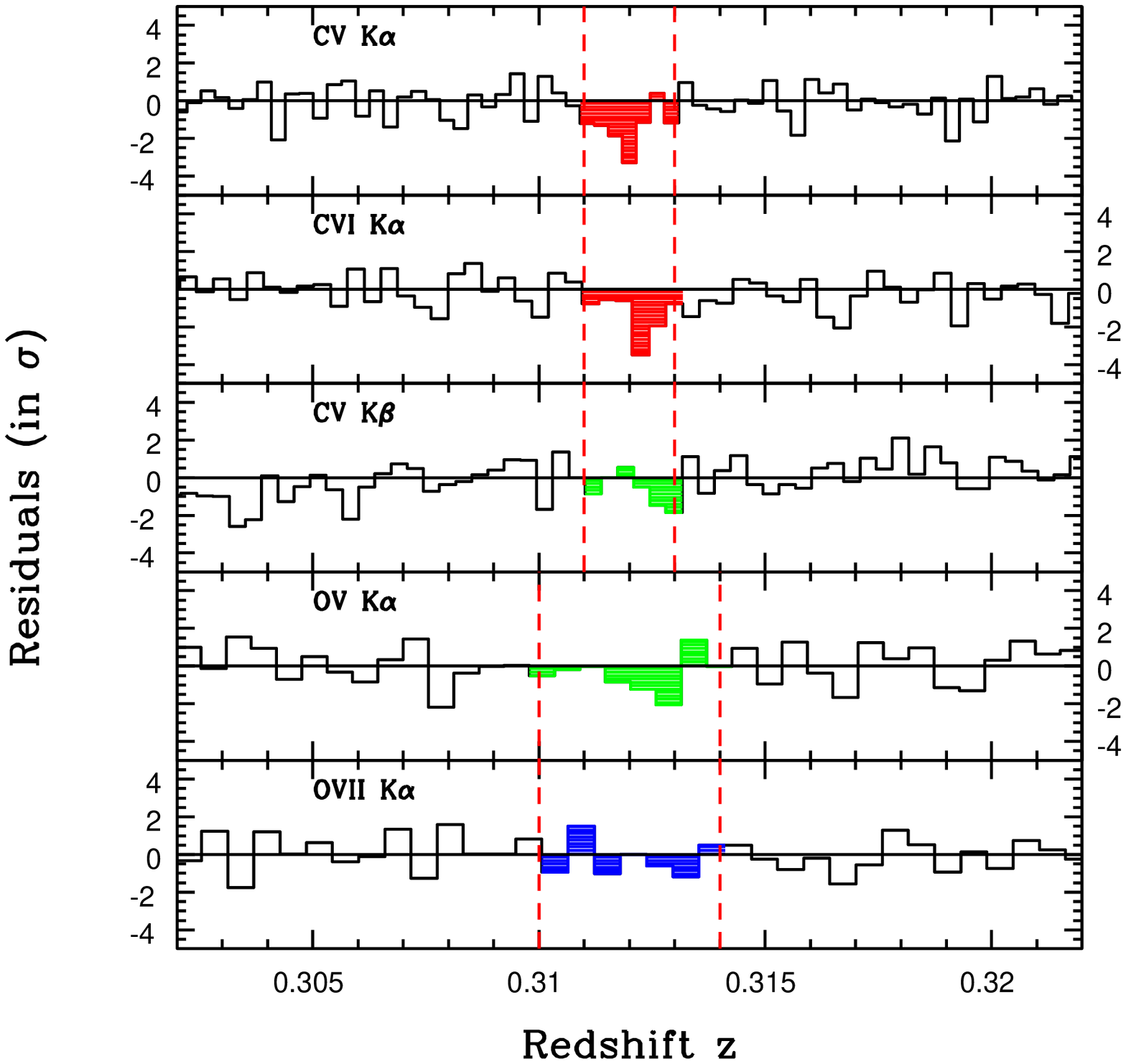,height=15.0cm,width=15.0cm,angle=0}
% un-comment the following line to include your fig1b.ps postscript file
%\psfig{figure=fig1b.ps,height=7.0cm,width=7.0cm,angle=-90}
}
\end{center}
\vspace{-1.5cm} 
\caption{\footnotesize Five portions of the LETG spectrum of 1ES~1553+113, centered around the X-ray redshift of the CV-CVI system, for 
CV K$\alpha$ (top panel), CVI K$\alpha$ (second panel), CV K$\beta$ (third panel), OV K$\alpha$, and OVII K$\alpha$. 
Detected lines are filled in red, lines hinted in the data are filled in green, while non detections have the spectral bins 
filled in blue. The two vertical dashed red lines delimit the X-ray redshift of the system: the wider interval for oxygen lines, compared to carbon lines, 
is due to the $\Delta\lambda = 40$ m\AA systematic unceratinty we assume throughout the entire LETG band, which translates into $\Delta z = 0.001$ 
and $\Delta z = 0.002$ at the wavelengths of the carbon and oxygen transitions, respectively.}
\label{fig3}
\end{figure}
%-----------------------------Figure End--------------------------------

\subsubsection{The $z_X = 0.237 \pm 0.001$ WHIM System}
The line at $\lambda = 49.80$ \AA\ is the second highest statistical significance line detected with the LETG (3.9$\sigma$).  
This line can only be associated with a COS absorber if identified as a CV K$\alpha$ at $z_X = 0.237 \pm 0.001$. 
Indeed, at a marginally consistent redshift of $z_{FUV} = 0.23559$, there is a weak (EW$=20 \pm 7$ m\AA) NLA ($b=12 \pm 8$ 
km s$^{-1}$) in the published COS spectrum (Danforth et al., 2010). 
The HI line, however, is too weak and much too narrow to be considered the actual counterpart of a putative CV absorber, so, if the LETG 
identification is correct, CV and NLA would be the tracers of a spatially co-located multi-phase WHIM filament, and a broader 
HI absorber should exist at $z\sim 0.237$, undetected down to the sensitivity limit of the published COS spectrum. 
To investigate this possibility further, we searched for such a BLA absorber in the currently available COS spectrum of 1ES~1553+113, which, 
at the interesting wavelengths has a factor of 2 higher S/N than the published spectrum. 
We confirm the presence of the NLA at $z_{FUV} = 0.23559$ with EW$=12 \pm 4$ m\AA\ and $b=34^{+17}_{-10}$ km s$^{-1}$ (Fig. 5). 
Additionally the data also show a structured and broad absorption complex at $z\simeq 0.2362 - 0.2378$ (Fig. 5). 
We model this complex with two gaussians with best-fitting HI Ly$\alpha$ centroids $z_{FUV}^1 = 0.23666$, $z_{FUV}^2 = 0.23734$, 
equivalent widths EW$_1 =25 \pm 5$ m\AA\ (5$\sigma$), EW$_2 = 13 \pm 6$ m\AA\ (2.2$\sigma$), and Doppler parameters b$_1 = 72 \pm 20$ 
km s$^{-1}$, b$_2 = 80^{+40}_{-20}$ km s$^{-1}$. The redshifts of these two BLAs are both fully consistent with that of the putative CV K$\alpha$ absorber, 
and their widths suggest temperatures in the ranges logT$_1 = 5.1 - 5.7$ and logT$_2 = 5.2 - 5.9$ (where the lower boundaries of these intervals are corrected for the average observed $b/b_{therm}=1.2$ ratio: Danforth et al., 2010), again consistent with the presence of a large fraction ($\sim 60-80$\%) of CV. 

%-----------------------------Figure Start------------------------------
\begin{figure}[h]
\begin{center}
\hbox{
% un-comment the following line to include your fig1a.ps postscript file
\hspace{1.0cm}
\psfig{figure=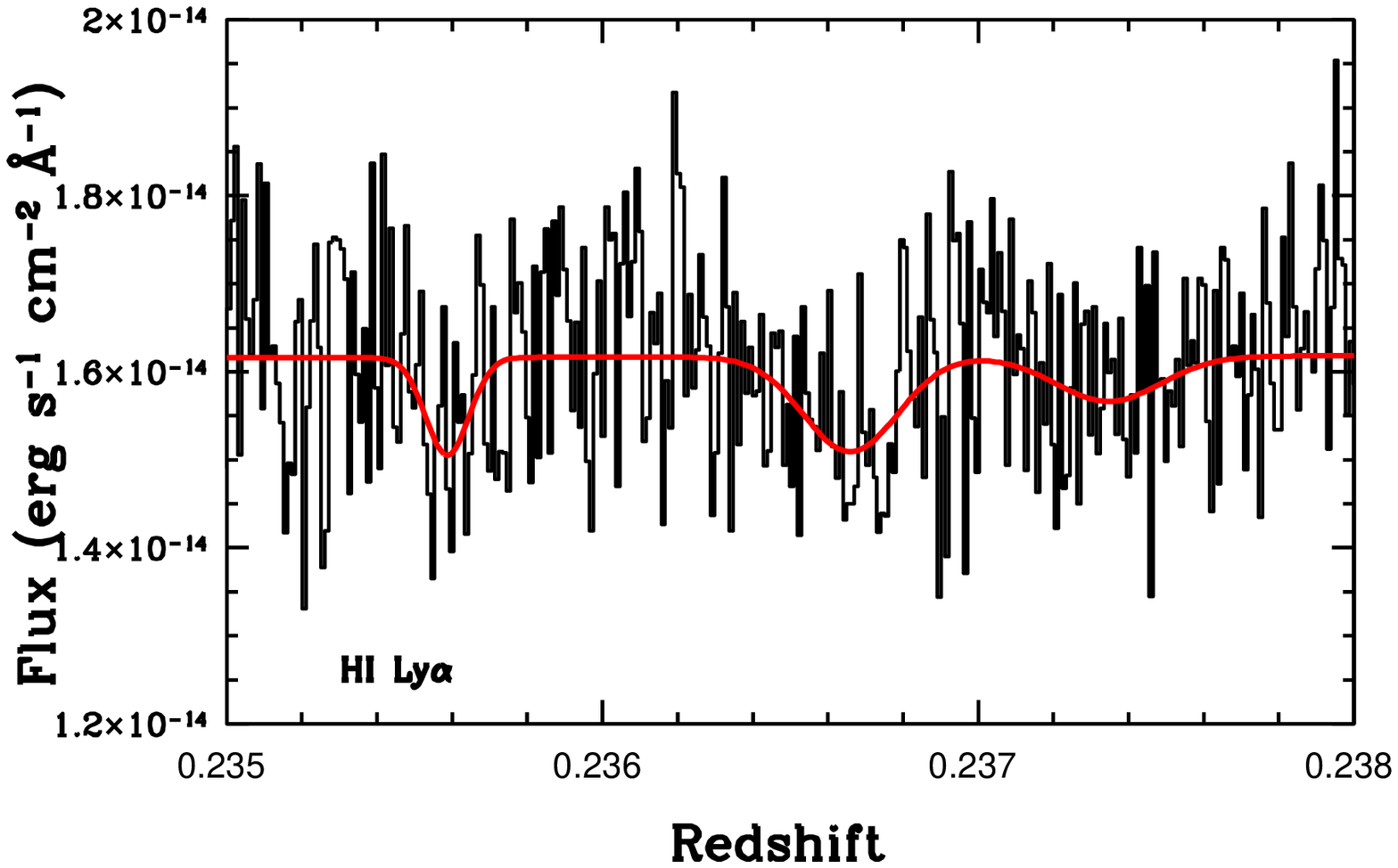,height=15.0cm,width=15.0cm,angle=0}
% un-comment the following line to include your fig1b.ps postscript file
%\psfig{figure=fig1b.ps,height=7.0cm,width=7.0cm,angle=-90}
}
\end{center}
\vspace{-1.5cm} 
\caption{\footnotesize Portion of the COS spectrum of 1ES~1553+113 where the HI Ly$\alpha$ transition at $z_{FUV} = 0.235-0.238$ lies. 
The red curve is the best-fitting continuum model plus three negative Gaussians modeling: (a) a NLA at $z_{FUV} = 0.23559$, (b) a BLA at 
$z_{FUV}^1 = 0.23666$, and (c) a BLA at $z_{FUV}^2 = 0.23734$}
\label{fig3}
\end{figure}
%-----------------------------Figure End--------------------------------

\noindent 
In the same temperature interval CVI is also relatively abundant, with a fraction raising monotonically from $f_{OVI} \simeq 15$\% at logT=5.1 to 
$f_{OVI} \simeq 50$\% at logT=5.9, but the oscillator strength of the CVI K$\alpha$ transition is 1.7 times smaller than that of the CV K$\alpha$ 
transition.  
The strongest transitions expected from oxygen, instead, depend on the exact temperature within this broad interval: at logT$\simeq 5.1 - 5.4$, 
OV K$\alpha$ is the strongest transition expected (with $f_{OV} \simeq 10-40$\%), while at logT$=\simeq 5.5-5.9$ the OVII K$\alpha$ 
dominates, with $f_{OVII} \simeq 70-90$\%. 
We therefore looked for the presence of these three transitions in the LETG spectrum of 1ES~1553+113. 
The top panel of Figure 6 shows the detection that we tentatively identify with CV K$\alpha$ at $z_X = 0.237$. The remaining three panels show 
the postions of the CVI K$\alpha$ (second panel), OV K$\alpha$ (third panel) and OVII K$\alpha$ (bottom panel) in the $z_X = 0.227-0.247$ redshift 
range. 
We note that OV K$\alpha$ and CVI K$\alpha$ (shifted by $\Delta z \simeq +0.001$) may be hinted in the data (second and third panels of Figure 6). 
We also note that the two line-like deficits of counts visible in the third and bottom panels of Figure 6 (magenta bins), just leftward of the 90\% 
confidence redshift interval of our tentative CV K$\alpha$ identification (dashed red lines), could be identifiable with OV K$\alpha$ and 
OVII K$\alpha$ at $z_X = 0.233 \pm 0.002$, respectively. Indeed the magenta-colored feature in the bottom panel of Fig. 6 is the lowest 
significance (1.7$\sigma$) line-like absorption feature identified by our scanning routine and susbequently confirmed by a Gaussian fit in the 
LETG, at $\lambda = 26.65$ \AA. However the redshift of these two putative OV K$\alpha$ and OVII K$\alpha$ identifications is only marginally 
(3$\sigma$) consistent with that of our CV K$\alpha$ identification. 
Moreover, for the putative OVII K$\alpha$ at $z=0.233 \pm 0.002$ an alternative identification is possible: that of an OV K$\alpha$ absorber at 
$z=0.191 \pm 0.002$ associated with a CV-BLA system (see \S 4.1.5). Higher quality data are needed to discriminate between these two possibilities. 

%-----------------------------Figure Start------------------------------
\begin{figure}[h]
\begin{center}
\hbox{
% un-comment the following line to include your fig1a.ps postscript file
\hspace{1.0cm}
\psfig{figure=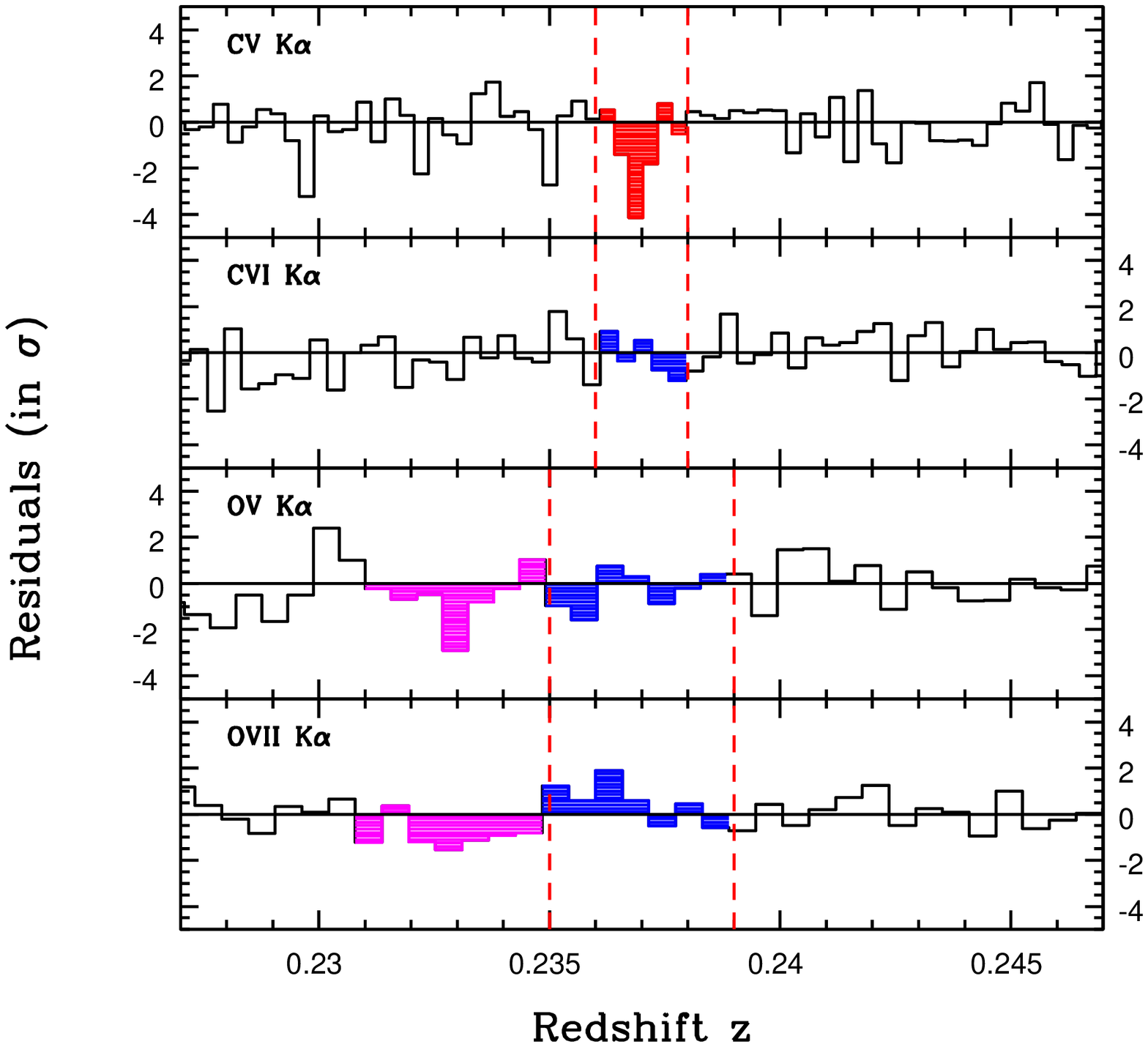,height=15.0cm,width=15.0cm,angle=0}
% un-comment the following line to include your fig1b.ps postscript file
%\psfig{figure=fig1b.ps,height=7.0cm,width=7.0cm,angle=-90}
}
\end{center}
\vspace{-1.5cm} 
\caption{\footnotesize Four portions of the LETG spectrum of 1ES~1553+113, centered around the X-ray redshift of the CV K$\alpha$ 
$z=0.237 \pm 0.001$ identification, for CV K$\alpha$ (top panel), CVI K$\alpha$ (second panel), OV K$\alpha$ (third panel), OVII K$\alpha$ (bottom panel). 
Detected lines are filled in red. Non detections have the spectral bins filled in blue. The two line-like features in the third and bottom panels, filled in 
magenta, are possible OV K$\alpha$ and OVII K$\alpha$ identifications at a redshift close to (but only marginally consistent with) that of the CV-BLA system. 
The two vertical dashed red lines delimit the X-ray redshift interval of the CV-BLA system.}
\label{fig3}
\end{figure}
%-----------------------------Figure End--------------------------------

We conclude that the most likely identification for the line at $\lambda = 49.80$ \AA\ (3.9$\sigma$) is that of a CV K$\alpha$ line imprinted 
by a highly ionized shock-heated WHIM metal system at $z_X = 0.237 \pm 0.001$, producing also structured BLA absorption at 
$z_{FUV}^1 = 0.23666$ and $z_{FUV}^2 = 0.23734$ (Fig. 5), as well as a nearby NLA at $z_{FUV} = 0.23559$ (Fig. 5) in a mildly photo-ionized portion of the 
filament, possibly at the interface of the shock-heated part of the main-body filament (Table 1; Fig. 1, blue lines; Fig. 5 and 6). 
Such multi-phase systems, with metals tracing hot collisionally ionized gas and narrow HI tracing a cooler photoionized portion of the same 
system, have been observed in OVI-NLA in the FUV (Danforth \& Shull 2008), but it would be the first time they are observed in hotter 
gas traced by CV-BLA pairs. 
Confirming the presence of a nearby higher-ionization OV-OVII absorber, would probe an even more extreme regime and would 
therefore be extremely important. 

\subsubsection{The Multiphase and Multi-Component $\langle z_X \rangle = 0.133 \pm 0.002$ CV-CVI-BLA System}
The line at $\lambda = 45.58$ \AA\ (3.8$\sigma$), cannot be associated with any metal or HI system in COS, if identified as CII, CIII or CVI. 
The CIV identification has already been ruled out in \S 4.1.2, because no corresponding CIV transition is detected in COS at $z=0.100 \pm 0.002$.
At the redshift of a putative CI identification ($z_X = 0.051$), instead, there is a NLA in COS ($z_{FUV} = 0.05094$). 
However, CI must be associated with almost neutral gas, with a large HI/H fraction. The HI column derived from the COS data is only 
N$_{HI} = 2\times 10^{13}$ cm$^{-2}$, implying a total baryon column of at most N$_H \sim 10^{15}$ cm$^{-2}$. 
Galactic CI is not detected in these LETG data, and our line of sight to 1ES1553+113 crosses a Galactic column density N$_H = 3.7 \times 10^{20}$ 
cm$^{-2}$. Finally, there are 10 CI transitions accessible in the COS bandpass, and none is detected at $z=0.051$. 
We therefore rule out the CI identification for this line. 

The only possibility left is CV at $z_X = 0.132 \pm 0.001$. 
The marginal detection at $\lambda = 38.25$ \AA\ (2.7$\sigma$) would have a marginally consistent redshift of $z_X = 0.134 \pm 0.001$, if 
identified with CVI K$\alpha$. 
At a redshift consistent with both the CV and CVI X-ray lines ($z_{FUV} = 0.13324$) there is a weak (N$_{HI} = 10^{13.15}$ cm$^{-2}$) HI Ly$\alpha$ in COS, 
with a doppler parameter of $b = 46 \pm 8$ km s$^{-1}$. Correcting for the empirically derived $b/b_{th}$ ratio of $\simeq 1.2$ (Danforth, Stocke 
\& Shull, 2010), gives a thermal width of $b_{th} = 38^{+16}_{-7}$ km s$^{-1}$, implying T $= (0.6 - 1.7) \times 10^5$ K, consistent with either 
shock-heated WHIM or low-density purely photoionized IGM. However, we note that, given the relatively low-temperature derived from the HI 
Doppler parameter, the column density of this HI component is probably too low to produce all the CV and CVI detected in the LETG, 
even assuming Solar abundances. Therefore, if our identification is correct, the CV-CVI system cannot be entirely and directly associated with 
the HI Ly$\alpha$ in COS. 
Possibly some of the detected CV and CVI must be associated with an even broader and shallower HI not detectable at the sensitivity of the current 
COS spectrum.  

To further investigate on this possible identification we visually inspected the portions of the LETG spectrum where the tentatively detected CV and 
CVI K$\alpha$ transitions, as well as other, similar ionization, associated lines of O and C would fall. 
Figure 7 shows 4 portions of the LETG spectrum of 1ES~1553+113 centered around the $z=0.131 - 0.135$ interval, for the CV K$\alpha$ (top panel), 
CVI K$\alpha$ (second panel), OIV K$\alpha$ (third panel) and OVII K$\alpha$ (bottom panel) transitions. 
Interestingly, the complex profiles of the CV and CVI K$\alpha$ lines, suggest the presence of a double component (red and green filled bins in the 
top and second panel of Fig. 7). No other associated X-ray line is hinted by the data (possibly with the exception of a small deficit of counts at one 
of the two components of OIV: third panel). 
 
%-----------------------------Figure Start------------------------------
\begin{figure}[h]
\begin{center}
\hbox{
% un-comment the following line to include your fig1a.ps postscript file
\hspace{1.0cm}
\psfig{figure=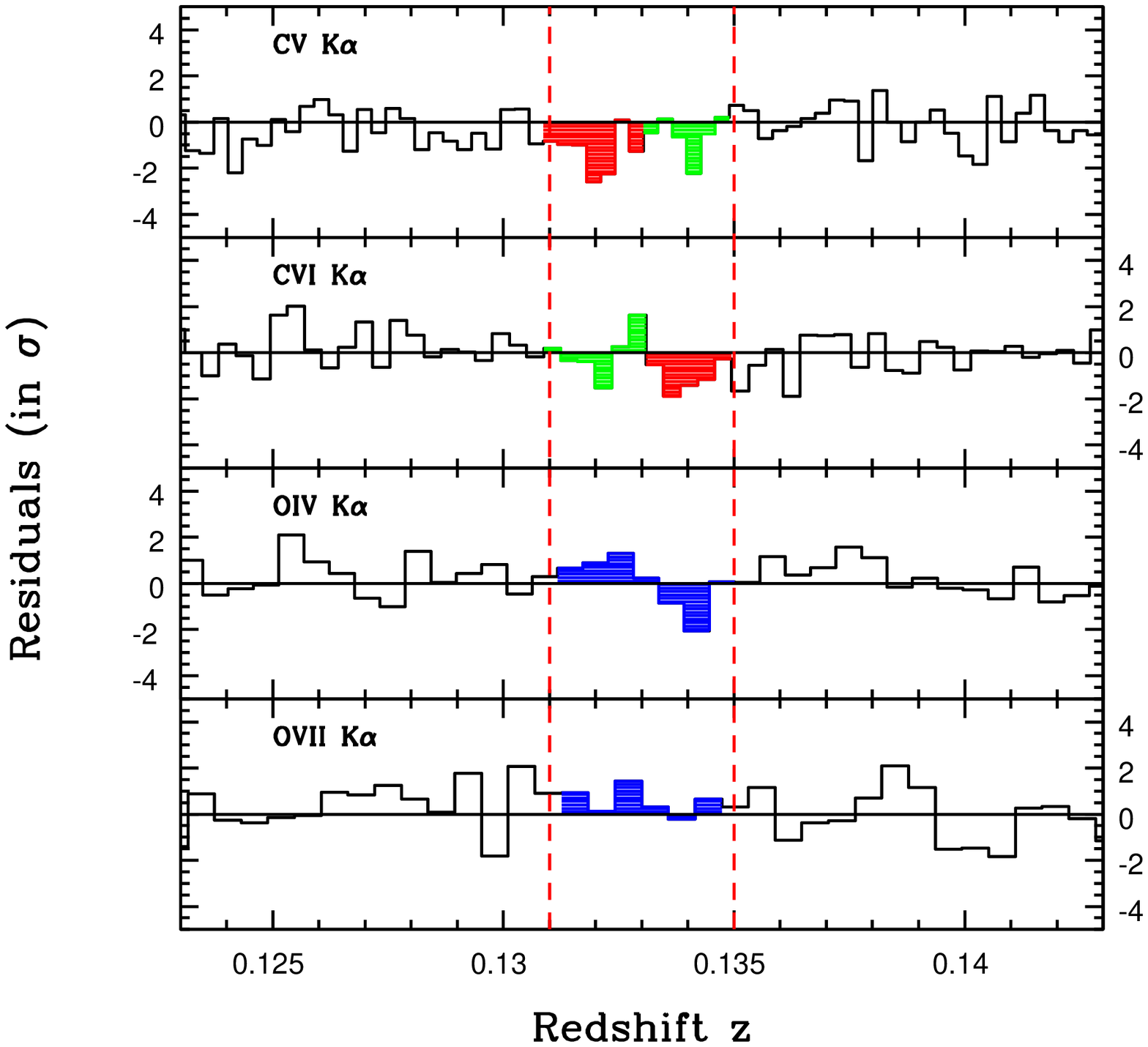,height=15.0cm,width=15.0cm,angle=0}
% un-comment the following line to include your fig1b.ps postscript file
%\psfig{figure=fig1b.ps,height=7.0cm,width=7.0cm,angle=-90}
}
\end{center}
\vspace{-1.5cm} 
\caption{\footnotesize Four portions of the LETG spectrum of 1ES~1553+113, centered around the X-ray redshift of the CV-CVI complex ($z=0.131 - 0.135$), 
for CV K$\alpha$ (top panel), CVI K$\alpha$ (second panel), OIV K$\alpha$ (third panel), OVII K$\alpha$ (bottom panel). 
Detected lines are filled in red, lines hinted in the data are filled in green, while non detections have the spectral bins 
filled in blue. The two vertical dashed red lines delimit the X-ray redshift interval of the system.}
\label{fig3}
\end{figure}
%-----------------------------Figure End--------------------------------
 
We conclude that the CV/CVI K$\alpha$ complex at an average redshift of $\langle z_X \rangle = 0.133 \pm 0.002$ (Fig. 7) 
is the most likely identification for these two LETG line complexes (Table 1; Fig. 1, green line, and Fig. 7 top 2 panels), possibly produced in a multi-phase, 
multi-component WHIM filament together with the spatially associated BLA absorber.

\subsubsection{The $z_X = 0.184-0.191$ Double-CV}
The line at $\lambda = 47.68$ \AA\ (3.6$\sigma$) can only be identified as CV K$\alpha$. 
No other ion of C can be associated with any metal and/or HI absorber in COS. 
Instead, the redshifts of the CV K$\alpha$ identification, $z_X = 0.184 \pm 0.001$ is close, but only marginally consistent, with the 
short-wavelength extreme of a narrow redshift interval, $z_{FUV} = 0.18640 - 0.18989$, where a rare triple-HI system is found in COS. 
No OVI is detected at $z_{FUV} = 0.18640$, while both the $z_{FUV} = 0.18773$ and the $z_{FUV} = 0.18989$ absorbers have associated 
OVI absorption, with the central system (the one with higher column density) showing also NV, CIII and SiII absorption (Danforth et al. 2010). 
We re-analyzed these three absorbers in the currently available HST-COS spectrum of 1ES~1553+113 and measured the doppler parameters 
of the lines detected from the two HI-metal systems at $z_{FUV} = 0.18773$ and $z_{FUV} = 0.18989$. From these, we derive the thermal HI Dopler 
parameters of these two systems (by correcting for the average $b/b_{therm}=1/2$ value): $b_{HI}^{therm} = 27 \pm 10$ km s$^{-1}$ and 
$b_{HI}^{therm} = 58 \pm 4$ km s$^{-1}$, respectively. Clearly, the HI-metal system at $z_{FUV} = 0.18773$ is imprinted by photo-ionized gas, 
while the $z_{FUV} = 0.18989$ HI-OVI absorber is a WHIM system with logT$=5.3 \pm 0.1$. 

Interestingly, in the LETG we also marginally detect (2.2$\sigma$) a line-like absorption feature at $\lambda = 47.94$ \AA. Also for this line, if real, the 
only possible identification, in association with a COS HI and/or metal system, is CV K$\alpha$ at $z_X = 0.191 \pm 0.001$, 
consistent with the long-wavelength extreme of the COS triple-HI-metal system, where a relatively strong BLA (logN$_{HI} = 13.565 \pm 0.017$) 
is present. 
Figure 8 shows five portions of the LETG spectrum of 1ES~1553+113 in the $z_X = 0.175 - 0.195$ interval, for the CV K$\alpha$ (top panel), the CVI K$\alpha$ 
(second panel), the OIV K$\alpha$ (third panel), the OV K$\alpha$ (fourth panel) and the OVII K$\alpha$ (bottom panel) transitions. 
No associated X-ray lines is hinted in the data for the tentative CV K$\alpha$ identification at $z=0.184 \pm 0.001$. The situation is different for 
the second tentative CV K$\alpha$ identification at $z=0.191 \pm 0.001$ (the higher redshift boundary of the triple-HI-metal system detected in COS, 
where a BLA is present). For this system the OV K$\alpha$ transition is identifiable with the lowest significance (1.7$\sigma$) line-like absorption feature 
identified by our scanning routine and susbequently confirmed by a Gaussian fit in the LETG, at $\lambda = 26.65$ \AA. 

%-----------------------------Figure Start------------------------------
\begin{figure}[h]
\begin{center}
\hbox{
% un-comment the following line to include your fig1a.ps postscript file
\hspace{1.0cm}
\psfig{figure=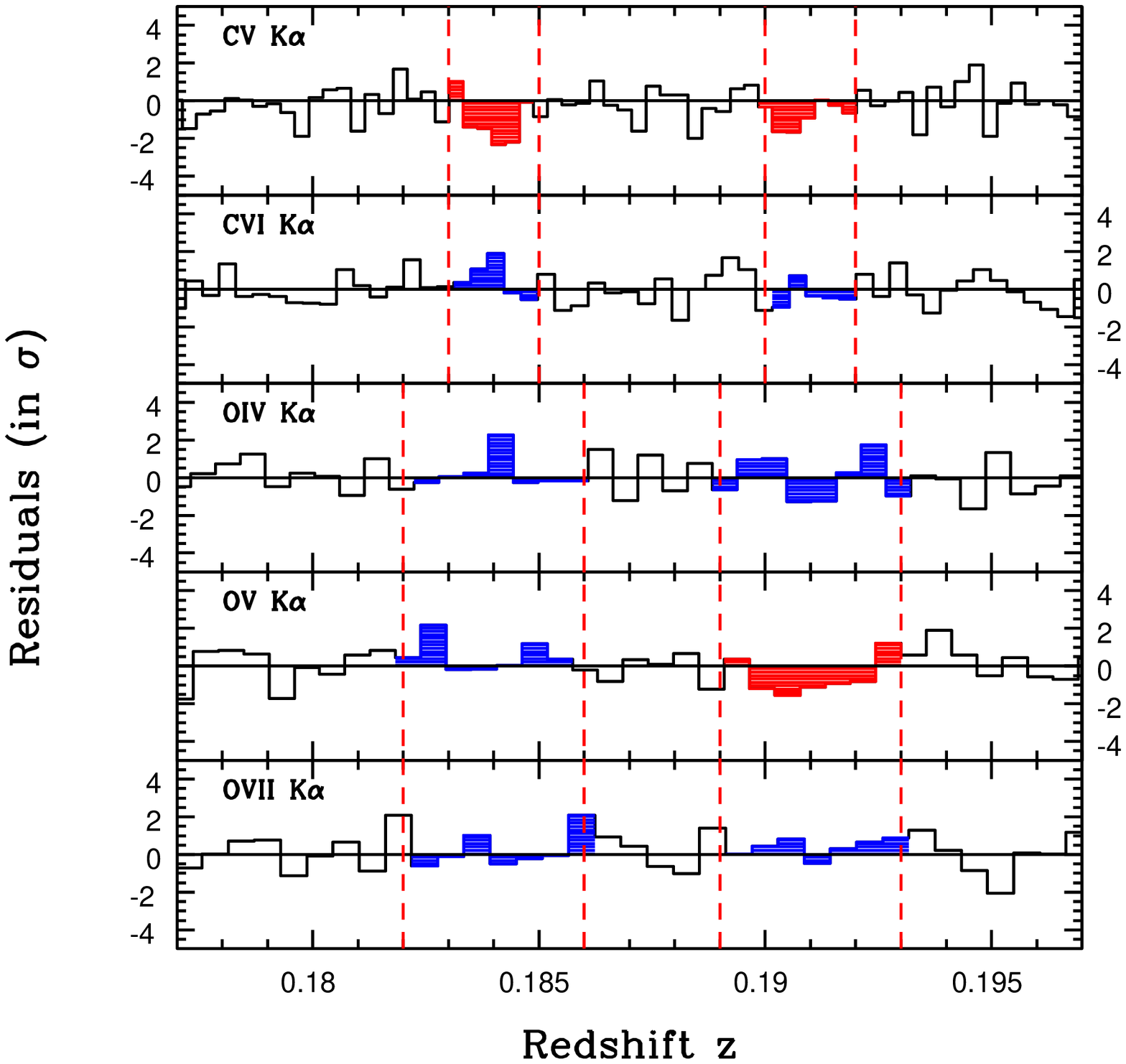,height=15.0cm,width=15.0cm,angle=0}
% un-comment the following line to include your fig1b.ps postscript file
%\psfig{figure=fig1b.ps,height=7.0cm,width=7.0cm,angle=-90}
}
\end{center}
\vspace{-1.5cm} 
\caption{\footnotesize Five portions of the LETG spectrum of 1ES~1553+113 in the $z_X = 0.175 - 0.195$ interval, for the CV K$\alpha$ (top panel), 
the CVI K$\alpha$ (second panel), the OIV K$\alpha$ (third panel), the OV K$\alpha$ (fourth panel) and the OVII K$\alpha$ (bottom panel) transitions. 
Detected lines are filled in red. Non detections have the spectral bins filled in blue. 
The two pairs of vertical dashed red lines delimit the X-ray redshift intervals of the CV K$\alpha$ identifications.}
\label{fig3}
\end{figure}
%-----------------------------Figure End--------------------------------
 
We thus speculate that these LETG lines at $\lambda = 47.68$ \AA\ (3.6$\sigma$), $\lambda = 47.94$ \AA (2.2$\sigma$) and $\lambda = 26.65$ \AA\ 
(only 1.7$\sigma$), may be the high-ionization CV K$\alpha$ ($z_X = 0.184$ and $z_X = 0.191$) and OV K$\alpha$ ($z_X = 0.191$) counterparts 
(Table 1; Fig. 8, top panel, red bins) of the interesting, likely multi-phase, triple-HI-metal system detected in COS, possibly witnessing the on-going 
interplay between nearby galaxy outflows and a WHIM filament at $z_{FUV} = 0.188989$, where a CV-OV-BLA system is present. 

\subsubsection{The $z_X = 0.041 \pm 0.02$ WHIM System}
Finally, the marginal detection at $\lambda = 22.47$ \AA\ ($2.3\sigma$) can only be redshifted OV or higher. 
The OV K$\alpha$ transition would be redshifted to $z_X=0.005 \pm 0.002$. 
There is a NLA present in COS at $z_{FUV} = 0.00717$, i.e. marginally consistent with the X-ray redshift. 
However, in photoionized gas OV shows fractional abundances similar to OVI. No OVI absorption is seen at 
this redshift either in the FUSE spectrum of 1ES~1553+113 or in its LETG spectrum. 
The OVI K$\alpha$ transition would be redshifted to $z_X = 0.020 \pm 0.002$. No HI Ly$\alpha$ or OVI$(1s^22s\rightarrow 1s^22p)$ 
is seen at a consistent redshifts in COS or FUSE. 
Finally, the OVII and OVIII K$\alpha$ transitions, would be redshifted to $z_X = 0.041 \pm 0.002$ and $z_X = 0.185 \pm 0.002$. 
These two redshifts are both consistent with either the broadest ($b_{HI} = 73 \pm 7$ km s$^{-1}$, corresponding to 
logT$ = 5.3 - 5.6$ K) and strongest (EW(BLA) = $135 \pm 14$ m\AA) BLA detected in the published COS spectrum, at $z_{FUV} = 0.04281$, 
or the redshift range of the unusual triple-HI-metal system at $z_{FUV}  = 0.18640 - 0.18989$, respectively. 
However, while a possible double-CV counterpart to the triple-HI-metal system seen in COS is possibly detected in the LETG (\S 4.1.5), no 
OVII counterpart is detected. So, it appears unlikely that the line at $\lambda = 22.47$ \AA\ is OVIII associated with the triple-HI-metal system. 

For this marginal LETG detection we therefore favor the $z_X = 0.041$ OVII-BLA association (Table 1, Columns 3-5; Fig. 1 magenta line), 
and tentatively identify this association with an intervening WHIM system at the COS-detected BLA signpost (though an OVIII-Triple-HI
association cannot be ruled out). No other associated X-ray line is hinted in the data (Fig. 9). 

%-----------------------------Figure Start------------------------------
\begin{figure}[h]
\begin{center}
\hbox{
% un-comment the following line to include your fig1a.ps postscript file
\hspace{1.0cm}
\psfig{figure=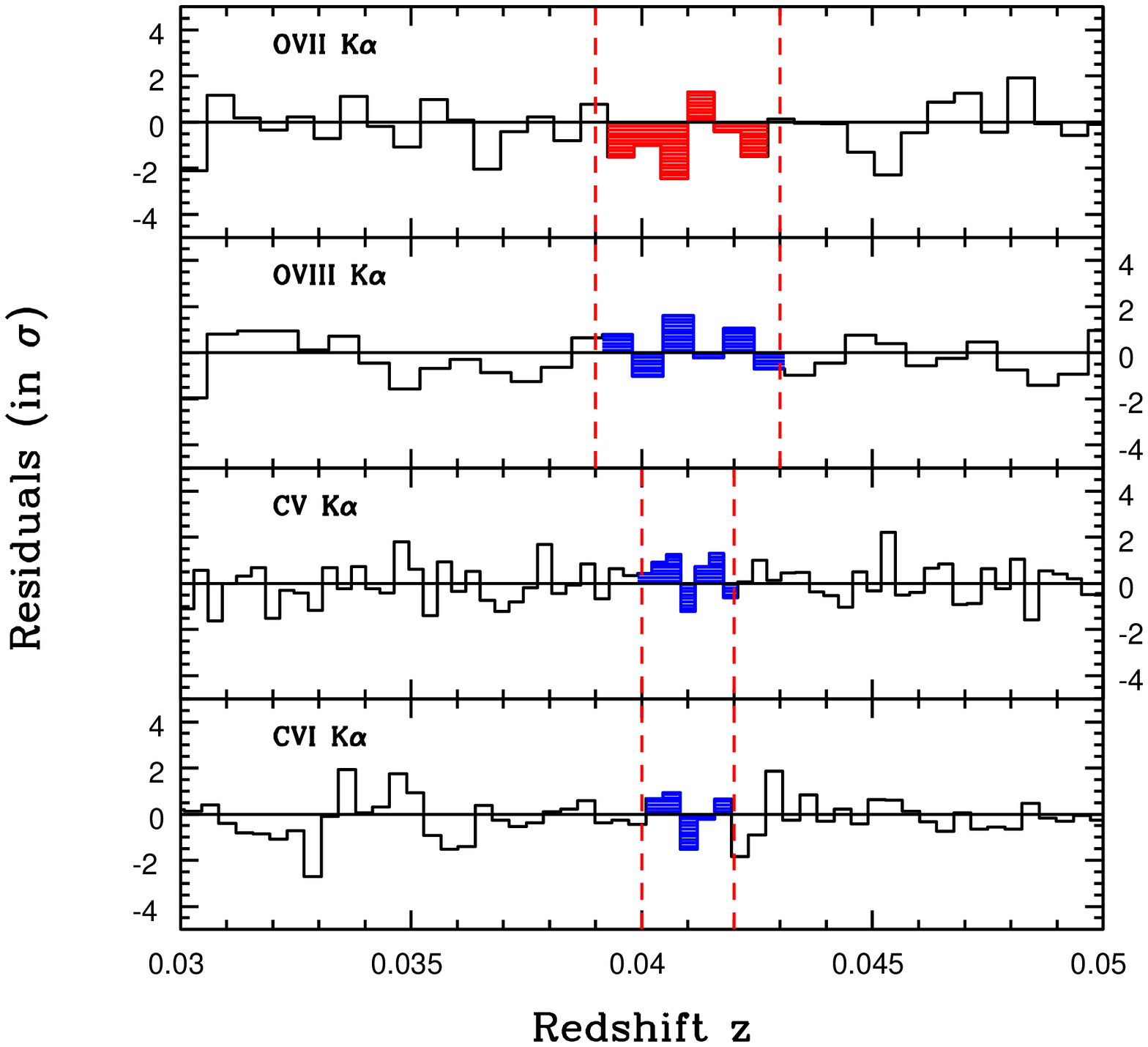,height=15.0cm,width=15.0cm,angle=0}
% un-comment the following line to include your fig1b.ps postscript file
%\psfig{figure=fig1b.ps,height=7.0cm,width=7.0cm,angle=-90}
}
\end{center}
\vspace{-1.5cm} 
\caption{\footnotesize Four portions of the LETG spectrum of 1ES~1553+113 in the $z_X = 0.03 - 0.05$ interval, for the OVII K$\alpha$ (top panel), 
the OVIII K$\alpha$ (second panel), the CV K$\alpha$ (third panel) and the CVI K$\alpha$ (bottom panel) transitions. 
Detected lines are filled in red. Non detections have the spectral bins filled in blue. 
The two vertical dashed red lines delimit the X-ray redshift intervals of the OVII K$\alpha$ identifications.}
\label{fig3}
\end{figure}
%-----------------------------Figure End--------------------------------

We stress that this is the weakest system for which we propose an identification in association with a COS system 
(only one LETG line, at $2.3\sigma$). However, given the nature of this association (OVII-BLA), in principle probing 
the majority of the baryons in the WHIM, a confirmation of this putative OVII K$\alpha$ detection at high statistical 
significance is highly desirable. 

\subsection{On the Statistical Significance of the X-ray Absorption Lines}
The statistiscal significances of the X-ray absorption lines reported in Table 1 (second column), are single-line statistical significances. 
That is the ratio between the line flux, and its 1$\sigma$ lower-bound uncertainty. 
For those lines not identifiable with $z \simeq 0 $ transitions imprinted by Galactic or circumgalactic gas, these significances 
over-estimate the true absorption-line statistical significances, because they derive from a number of redshift trials. 
This number of redshift trials, depends on the wavelength of the transition, the spectrometer resolution, and the available path length 
over which the blind search is performed: N$= (\lambda_0 \times \Delta z) / \Delta\lambda_{spec}$, where $\lambda_0$ is the rest frame 
wavelength of the given transition, and $\Delta\lambda_{spec}$ the spectrometer LSF FWHM. 
In our case, $\Delta z = 0.4$ and $\Delta\lambda_{spec} = 50$ m\AA. So, the number of independent LETG resolution elements  is 173 for 
the OVII K$\alpha$ and 322 for the CV K$\alpha$. 
In such conditions, the true statistical significance (i.e. after correcting for the unknown line position) of a line blindly detected at a 
``single-line'' statistical significance of 4$\sigma$ is only 2.6$\sigma$ and 2.4$\sigma$, for OVII K$\alpha$ and CV K$\alpha$ respectively.

However, if an independent (and physically self-consistent) redshift prior is known (OVI or BLA in COS), then the line identification can be 
considered ``secure'' and therefore the single-line statistical significance provides a reliable estimate of the true statistical significance 
of the line. 
This is true for five out of the eight intervening lines: (a) the highest statistical significance $z_X = 0.312$ CV and CVI lines, at redshift consistent with 
that of a strong BLA-OVI system with the right temperature to produce CV, (b) the $z_X = 0.237$ CV line, at redshift consistent 
with that of a shallow BLA complex with the right temperature to produce CV; (c) the putative $z_X = 0.191$ CV-OV lines, at redshifts 
consistent with that of a BLA at $z_{FUV} = 0.18989$ in COS, and (d) the speculative $z_X = 0.041$ OVII absorber, at redshift consistent with that 
of a strong BLA whose thermal width is appropriate for the production of OVII. 

The remaining three lines are identified as CV K$\alpha$ at $z_X = 0.184$ (close to the lower boundary of a rare triple HI-metal system detected in 
COS), and CV-CVI K$\alpha$ at $\langle z_X \rangle = 0.133$, a redshift consistent with that of a BLA whose column density, however, is 
probably too low to produce the entire CV and CVI absorption (given the relatively low temperature derived for this BLA). 
For the putative CV K$\alpha$ line at $z_X = 0.184 \pm 0.001$, its single-line significance is 3.6$\sigma$, which corresponds 
to a probability of chance detection of 3.5$\times 10^{-4}$. Allowing for the number of redshift trials, the probability of chance detection 
of this line rises to 0.056, a Gaussian-equivalent significance of 1.9$\sigma$. 
Fo the CV-CVI system at $\langle z_X \rangle = 0.133$, the presence of two lines poses a bound on the number of redshift trials for 
the line detected at lower statistical significance and makes its measured single-line statistical significances close to the true 
statistical significances. These two lines are detected at single-line significances of 3.8$\sigma$ and $2.7\sigma$, corresponding to 
probabilities of chance detections of 1.5$\times 10^{-4}$ and $7\times 10^{-3}$, respectively. 
Allowing for the number of redshift trials, the probability of chance detection of the strongest of these two lines rises to 0.024, 
a Gaussian-equivalent significance of 2.3$\sigma$. 
However, considering this as a prior for the second line, the probability of having by chance both lines (independently) is the product 
of the two probabilities: P$ = 0.024 \times 0.007 = 1.7\times 10^{-4}$, corresponding to a Gaussian-equivalent significance 
of 3.8$\sigma$. 

\section{Conclusions}
We report on the detection of six absorption lines and five additional features associated with either other X-ray lines detected at higher significance 
and/or Far-Ultraviolet (FUV) signposts, in the 500 ks {\em Chandra}-LETG spectrum of the bright (F$_{0.1-2} = 1.2$ mCrab) 
and distant ($z\ge 0.4$) blazar 1ES~1553+113. 
The six detections have single-line statistical significances between $3.6-4.1\sigma$, while the five additional associations are detected with 
single-line statistical significances of 2.2-2.8$\sigma$. 

We tentatively identify these absorption lines as belonging to six different systems: the Galaxy cold ISM, and five intervening warm-hot 
IGM filaments.

Three out of the five intervening absorption systems show multiple-ion associations in the X-rays, have BLA (and OVI for the system 
at $z_X = 0.312$) counterparts at consistent redshifts in the {\em HST} COS spectrum of 1ES~1553+113 and have true X-ray statistical 
significances of $5.8\sigma$ ($z_X = 0.312$; 6.3$\sigma$ if the low-significance OV and CV K$\beta$ associations are considered), 
$3.9\sigma$ ($z_X = 0.237$) and 3.8$\sigma$ ($\langle z_X \rangle = 0.133$). 
For these three systems, the COS counterparts are also physically consistent with the ionization degree and temperature of the X-ray absorbers in a 
shock-heated gas scenario. However, for the $\langle z_X \rangle = 0.133$ system, given the temperature implied by the BLA width, the HI 
column density is probably too low to produce all the detected CV and CVI absorption.  

The two remaining proposed identifications are X-ray single-ion tentative detections (with the exception of the CV-BLA-OVI $z_X = 0.191$ system, 
where OV may also be present), both with a possible, physically consistent, COS counterpart: (b) the double CV K$\alpha$ system at $z_X = 
0.184$ (3.6$\sigma$) and $z_X = 0.191$ (2.2$\sigma$, plus OV K$\alpha$, 1.7$\sigma$), both with HI (BLA-OVI for the putative CV-OV at 
$z_x = 0.191$) counterparts, and (c) the possible OVII K$\alpha$ absorber at $z_X = 0.041$ (2.3$\sigma$) with a BLA counterpart. 

Here we summarize our identifications: 

\begin{itemize}
\item{Galactic ISM:} 
We detect atomic OI (and possibly an excess of ISM oxygen in compound forms, over the instrumental molecular OI), as well as CII at $z \simeq 0$ 
and associate these absorbers with the cold ISM of our Galaxy along the line of sight to 1ES~1553+113. 

\item{Intervening System I:}
this is the highest significance LETG system, with two lines detected at single-line statistical significances of 4.1$\sigma$ each. 
We identify these two lines with CV and CVI absorption at $z_X=0.312 \pm 0.001$, possibly imprinted by an intervening WHIM filament, as further 
suggested by our a-posteriori discovery of an associated strong BLA and OVI counterpart in the COS spectrum, with the right temperature 
(i.e. line Doppler parameters) and strength. Associated OV and CV K$\beta$ absoprtion are also possibly hinted by the X-ray data. 
The combined (in quadrature) statistical significance of this system in the LETG (CV and CV K$\alpha$ lines only)is 5.8$\sigma$. 
If the association with OV K$\alpha$ and CVI K$\beta$ is considered, the statistical significance of the system is 6.3$\sigma$).

\item{Intervening System II:}
we identify CV absorption at $z_X = 0.237 \pm 0.001$, consistent with the redshift of a structured 
BLA complex that we discovered a posteriori in the currently avaiable COS spectrum of 1ES~1553+113. 
Given the presence of a physically self-consistent redshift prior (the BLA complex in COS), the true statistical significance 
of this line in the LETG is of 3.9$\sigma$.

\item{Intervening System III:}
we identify a CV-CVI-BLA system at $\langle z_X \rangle = 0.133 \pm 0.002$, and associate this absorber with an intervening warm 
WHIM filament, where only part of the X-ray absorption is physically related to the detected BLA. 
The statistical significance of this system in the LETG (after properly accounting for the number of redshift trials) is 3.8$\sigma$.

\item{Tentative Intervening System IV:}
we identify 2 CV K$\alpha$ absorbers at redshifts $z_X = 0.184 \pm 0.001$ and $z_X = 0.191 \pm 0.001$, where also OV K$\alpha$ is 
tentatively detected, adjacent to the redshifts of the extremes of a narrow redshift interval containing a rare triple-HI-metal absorption 
system. One of these three HI absorber is a BLA-OVI pair, with redshift and temperature fully consistent with that of the CV-OV absorber 
at $z_X = 0.191 \pm 0.001$. 
We speculate that the X-ray CV-OV absorbers trace the same multi-phase medium traced by the triple-HI-metal system in COS, 
possibly representing the result of the interplay between galaxy winds and a WHIM filament at $z_X = 0.191 \pm 0.001$. 
The CV line at $z_X = 0.184 \pm 0.001$ is detected at single-line statistical significance of 3.6$\sigma$ (a true statistical significance 
of only 1.9$\sigma$, after allowing for the number of redshift trials), while the CV-BLA-OVI system at $z_X = 0.191 \pm 0.001$ is detected 
at a true significane of and 2.2$\sigma$ (2.8$\sigma$ if the OV K$\alpha$ association is considered).

\item{Tentative Intervening System V:}
we tentatively identify an OVII-BLA system at $z_X=0.041 \pm 0.002$ and associate this absorber with an intervening hot WHIM filament. 
This is the weakest system for which we report a possible identification, with only one X-ray line at a single-line statistical significance 
of 2.3$\sigma$, but with a possible redshift prior identified in the  broadest and strongest BLA reported in the published COS spectrum. 

\end{itemize}

This moderate quality 500 ks {\em Chandra}-LETG spectrum of the, by far, brightest steady soft X-ray target in the $z > 0.4$ sky, 
strongly hints at the existence of possible X-ray metal WHIM counterparts to genuine gaseous signposts, previously 
secured with the {\em HST}-COS (or viceversa), and so demonstrates the efficiency of this observational strategy and 
the power of the X-ray-FUV synergy for WHIM studies.
Moreover, these results suggest that the gas may be often found to be multiphase, possibly indicating that the IGM structure  is richer 
than typically found in hydrodynamical simulations.
Longer integrations with both the high resolution soft X-ray spectrometers available (the {\em Chandra}-LETG and 
the XMM-{\em Newton}-RGSs) are needed to confirm at high statistical significances and with two indipendent measurements 
these tentative identifications and associations (especially for the three lowest statistical-significance associations), and to conduct 
a sensitive exploration of even hotter WHIM systems along this line of sight. 

\section{Acknowledgements}
We thank Ehud Behar for providing wavelength, oscillator strengths and transition probabilities of inner shell transitions 
of C, O and Ne. 
FN acknowledges support from CXO grant GO1-12174X, and from INAF-ASI ADAE grant 1.05.04.13.15. 
MS and CD 
%, JS and YY 
acknowledge support from COS grant NNX08AC14G to the University of Colorado, from NASA. 
XB acknowledges support by the Spanish Ministry of Economy and Competitiveness through grant AYA2010-21490-C02-01.

%%%%%%%%%%%%%%%%%%%%%%%%%%%%%%%%%%%%%%%%%%%%%%%%%%%%%%

%%%%%%%%%%%%%%%%%%

\begin{references}
Asplund, M., Grevesse, N., Sauval, A.J., Scott, P., 2009, ARA\&A, 47, 481 \\
Bar-Shalom, A., Klapish, M. \& Oreg, J., JQSRT, 71, 169 \\
Behar, E. \& Netzer, H., 2002, ApJ, 570, 165 \\
Bennett, C.L. et al., 2003, ApJS, 148, 1 \\
Bertone S., Schaye J., Dalla Vecchia C., Booth C. M., Theuns T., Wiersma R. P. C., 2010, MNRAS, 407, 544 \\
Branchini, E., et al., 2009, ApJ, 697, 328 \\
Brinkman, A.C., et al., 2000, ApJ, 530, L111B \\ 
Buote, D.A., Zappacosta, L., Fang, T., Humphrey, P.J., Gastaldello, F., Tagliaferri, G., 2009, ApJ, 695, 1351 \\  
Cen, R. \& Ostriker, J.P., 2006, ApJ, 650, 560 \\
Cen, R. \& Fang, T., 2006, ApJ, 650, 573 \\
Danforth, C.W. et al., 2011, ApJ, 743, 18 \\
Danforth, C.W. et al., 2010, ApJ, 720, 976 \\
Danforth, C.W., Stocke, J.T. \& Shull, J.M., 2010, ApJ, 720, 976 \\  
Danforth, C.W. \& Shull, J.M., 2008, ApJ, 679, 194 \\  
Fang, t. et al., 2010, ApJ, 714, 1715 \\
Fang, T., Canizares, C.R., \& Yao, Y., 2007, ApJ, 670, 992 \\ 
Gupta, A. et al., 2012, ApJ, 756, L8 \\
den Herder, J.W. et al., 2001, A\&A 365, L7 \\ 
Juett, A.M., Schulz, N.S. \& Chakrabarty, D., 2004, ApJ, 612, 308 \\
Kaastra, j.S. et al., 2011, A\&A, 534, 37 \\
Kaastra, J.S., Werner, N., den Herder, J.W.A., Paerels, F.B.S., de Plaa, J., Rasmussen, A.P., de Vries, C.P., 
2006, ApJ, 652, 189 \\ 
Kalberla, P.M.W. et al., 2005, A\&A, 440, 775 \\
Keshet, U., Kushnir, D., Loeb, A., Waman, E., 2012, arXiv:1210.1574 \\
Kirkman, D. et al., 2003, ApJS, 149, 1 \\
Mathur, S., Weinberg, D.H. \& Chen, X., 2003, ApJ, 582, 82 \\
Nicastro, F. et al., 2010, ApJ, 715, 854 \\
Nicastro, F., Mathur, S. \& Elvis, M., 2008, Science, 319, 55 \\ 
Nicastro, F. et al., 2005a, Nature, 433, 495 \\
Nicastro, F. et al., 2005b, ApJ, 629, 700 \\ 
Nicastro, F. et al., 2002, ApJ, 573, 157 \\
Murray, S.S. \& Chappell, J.H., 1995, SPIE, 597, 274 \\ 
Rasmussen, A.P., Kahn, S.M., Paerels, F., den Herder, J.W., Kaastra, J., de Vries, C., 2007, ApJ, 656, 129 \\  
Rauch, M., 1998, ARA\&A, 36, 267 \\
Schulz, N.S., Juett, A., Chakrabarty, D., Canizares, C.R., 2003, AN, 324, 166 \\
Shull, J.M., Smith, B.D., Danforth, C.W., 2012, ApJ, 759, 23 \\
Smith, B. D,  Hallman E. J., Shull J. M., O'Shea B. W., 2011, ApJ, 731, 6 \\
Spergel, D.N. et al., 2007, ApJS, 170, 377 \\
Takei, Y., Fujimoto, R. \& Mitsuda, K., 2002, ApJ, 581, 307 \\
Tornatore L., Borgani S., Viel M., Springel V., 2010, MNRAS, 402, 1911 \\
Treves, A., Falomo, R. \& Uslenghi, M., 2007, A\&A, 473, L17 \\
Verner, D.A., Verner, E.M. \& Ferland, G.J., 1996, Atom.Ph, 4003 \\
Weinberg, D.H., 1997, ApJ, 490, 564 \\
Williams, R.J., Mathur, S., Nicastro, F., Elvis, M., 2006, ApJ, 624, L95 \\
Williams, R.J, Mulchaey, J.S., Kollmeier, J.A. \& Cox, T.J., 2010, ApJ, 724, L25 \\  
Yao, Y., Shull, J.M.,, Wang, Q.D., Cash, W., 2012, ApJ, 746, 166 \\
Yoshikawa, K. et al., 2003, PASJ, 55, 879 \\
Zappacosta, L., Nicastro, F., Krongold, Y., Maiolini, R., 2012, ApJ, 753, 137 \\
Zappacosta, L. et al., 2010, ApJ, 717, 74 \\
\end{references}
\end{document}